\documentclass[a4paper,fleqn,usenatbib]{mnras}

\usepackage[T1]{fontenc}
\usepackage{ae,aecompl}
\usepackage{newtxtext,newtxmath}

\usepackage{graphicx}
\usepackage{amsmath}	
\usepackage{amssymb}	

\usepackage{epstopdf}

\title[Radial velocities of solar neighborhood RR Lyraes]{A new method of 
measuring center-of-mass velocities of radially pulsating stars from high-resolution 
spectroscopy. \thanks{Based on data collected with UVES@VLT under program ID
083.B-0281 and with SARG@TNG under program IDs AOT~19 TAC~11 and AOT~20 TAC~83.
Also based on ESO FEROS and HARPS archival reduced data products, under program
IDs 079.D-0462 and 178.D-0361. Also based on data of RR~Lyrae obtained with the
2.7m telescope at the McDonald Observatory, TX, USA.}}

\author[N. Britavskiy et al.]
{N. Britavskiy$^{1, 2, 3}$\thanks{email:britvavskiy@gmail.com},
E. Pancino$^{4,5}$,
V. Tsymbal$^{6}$,
D. Romano$^{7}$,
and L. Fossati$^{8}$\\
$^{1}$Instituto de Astrofisica de Canarias, E-38205 La Laguna, Tenerife, Spain\\
$^{2}$Universidad de La Laguna (ULL), Dpto. de Astrofi­sica, E-38206 La Laguna, Tenerife, Spain\\
$^{3}$IAASARS, National Observatory of Athens, GR-15236 Penteli, Greece\\
$^{4}$INAF - Osservatorio Astrofisico di Arcetri, Largo Enrico Fermi 5, 50125 Firenze, Italy\\
$^{5}$ASI Science Data Center, Via del Politecnico snc, 01333 Roma, Italy\\
$^{6}$Crimean Federal University, 295007, Vernadsky av. 4, Simferopol, Crimea\\
$^{7}$INAF-Osservatorio Astronomico di Bologna, Via Gobetti 93/3, I-40129 Bologna, Italy\\
$^{8}$Space Research Institute, Austrian Academy of Sciences, Schmiedlstrasse 6, A-8042 Graz, Austria}
\begin{document}

\date{Accepted 2017 November 10. Received 2017 November 09; in original form 2017 July 14.}

\pagerange{\pageref{firstpage}--\pageref{lastpage}} \pubyear{2017}

\maketitle
\label{firstpage}

\begin{abstract} We present a radial velocity analysis of 20 solar neighborhood RR Lyrae and 3 Population II Cepheids variables. We obtained high-resolution, moderate-to-high signal-to-noise ratio spectra for most stars and obtained spectra were covering different pulsation phases for each star. To estimate the gamma (center-of-mass) velocities of the program stars, we use two independent methods. The first, `classic' method is based on RR Lyrae radial velocity curve templates. The second method is based on the analysis of absorption line profile asymmetry to determine both the pulsational and the gamma velocities. This second method is based on the Least Squares Deconvolution (LSD) technique applied to analyze the line asymmetry that occurs in the spectra. We obtain measurements of the pulsation component of the radial velocity with an accuracy of $\pm$ 3.5 km~s$^{-1}$. The gamma velocity was determined with an accuracy $\pm$ 10 km~s$^{-1}$, even for those stars having a small number of spectra. The main advantage of this method is the possibility to get the estimation of gamma velocity even from one spectroscopic observation with uncertain pulsation phase. A detailed investigation of the LSD profile asymmetry shows that the projection factor $p$ varies as a function of the pulsation phase -- this is a key parameter which converts observed spectral line radial velocity variations into photospheric pulsation velocities. As a byproduct of our study, we present 41 densely-spaced synthetic grids of LSD profile bisectors that are based on atmospheric models of RR Lyr covering all pulsation phases.
\end{abstract}


\section{Introduction}
\begin{table*}
\caption{Basic information for the program stars.}
\label{tab:info}
\begin{tabular}{lcccccll}
\hline
Star     & R.A.(J2000) & Decl.(J2000)  & Type &  $V$	  & $\Delta V$  & Epoch	    & Period   \\
         & (h m s) & ($\degr$ $\arcmin$ $\arcsec$) & & (mag)   & (mag)  & (JD 2400000$+$)  & (day)\\
\hline
DR And   & 01 05 10.71 & $+$34 13 06.3 & RRab *&  11.65 -- 12.94 &1.29 & 51453.158583 & 0.5631300 \\
X Ari    & 03 08 30.88 & $+$10 26 45.2 & RRab &  11.28 -- 12.60 &1.32 & 54107.2779   & 0.6511681 \\
TW Boo   & 14 45 05.94 & $+$41 01 44.1 & RRab &  10.63 -- 11.68 &1.05 & 53918.4570   & 0.53226977\\
TW Cap   & 20 14 28.42 & $-$13 50 07.9 & CWa  &   9.95 -- 11.28 &1.33 & 51450.139016 & 28.610100 \\
RX Cet   & 00 33 38.28 & $-$15 29 14.9 & RRab *&  11.01 -- 11.75 &0.74 & 52172.1923   & 0.5736856 \\
U Com    & 12 40 03.20 & $+$27 29 56.1 & RRc  &  11.50 -- 11.97 &0.47 & 51608.348633 & 0.2927382 \\
RV CrB   & 16 19 25.85 & $+$29 42 47.6 & RRc  &  11.14 -- 11.70 &0.56 & 51278.225393 & 0.3315650 \\
UZ CVn   & 12 30 27.70 & $+$40 30 31.9 & RRab &  11.30 -- 12.00 &0.70 & 51549.365683 & 0.6977829 \\
SW CVn   & 12 40 55.03 & $+$37 05 06.6 & RRab &  12.03 -- 13.44 &1.41 & 51307.226553 & 0.4416567 \\
AE Dra   & 18 27 06.63 & $+$55 29 32.8 & RRab &  12.40 -- 13.38 &0.98 & 51336.369463 & 0.6026728 \\
BK Eri   & 02 49 55.88 & $-$01 25 12.9 & RRab &  12.00 -- 13.05 &1.05 & 51462.198773 & 0.5481494 \\
UY Eri   & 03 13 39.13 & $-$10 26 32.4 & CWb  &  10.93 -- 11.66 &0.73 & 51497.232193 & 2.2132350 \\
SZ Gem   & 07 53 43.45 & $+$19 16 23.9 & RRab &  10.98 -- 12.25 &1.27 & 51600.336523 & 0.5011365 \\
VX Her   & 16 30 40.80 & $+$18 22 00.6 & RRab *&   9.89 -- 11.21 &1.32 & 53919.451    & 0.45536088\\
DH Hya   & 09 00 14.83 & $-$09 46 44.1 & RRab &  11.36 -- 12.65 &1.29 & 51526.426583 & 0.4889982 \\
V Ind    & 21 11 29.91 & $-$45 04 28.4 & RRab &   9.12 -- 10.48 &1.36 & 47812.668    & 0.479601  \\
SS Leo   & 11 33 54.50 & $-$00 02 00.0 & RRab &  10.38 -- 11.56 &1.18 & 53050.565    & 0.626335  \\
V716 Oph & 16 30 49.47 & $-$05 30 19.5 & CWb  &   8.97 --  9.95 &0.98 & 51306.272953 & 1.1159157 \\
VW Scl   & 01 18 14.97 & $-$39 12 44.9 & RRab &  10.40 -- 11.40 &1.00 & 27809.381    & 0.5109147 \\
BK Tuc   & 23 29 33.33 & $-$72 32 40.0 & RRab &  12.40 -- 13.30 &0.90 & 36735.605    & 0.5502000 \\
TU UMa   & 11 29 48.49 & $+$30 04 02.4 & RRab &   9.26 -- 10.24 &0.98 & 51629.148846 & 0.5576587 \\
RV UMa   & 13 33 18.09 & $+$53 59 14.6 & RRab &   9.81 -- 11.30 &1.49 & 51335.380433 & 0.4680600 \\
UV Vir   & 12 21 16.74 & $+$00 22 03.0 & RRab *&  11.35 -- 12.35 &1.00 & 51579.459853 & 0.5870824 \\
\hline
\end{tabular}
\begin{flushleft}
{\em Notes.} The coordinates (columns 2 and 3), variability class and stars with exhibiting the Blazhko effect \citep[according to][]{darta_blazhko} are marked by an asterisk (column 4), magnitude ranges (column 5), amplitude of the magnitude variation (column 6), and pulsation periods
(column 8) have been taken from the General Catalog of Variable Stars
\citep[GCVS,][]{samus10}. The epochs of maximum light (column 7) were extracted from the
ROTSE light curves (see Paper~I for more details).\end{flushleft}
\end{table*}

RR Lyraes are important tracers of galactic dynamics and evolution. Their high
luminosity makes them good tracers for investigations of the Galactic halo
\citep[e.g.][]{koll09} and of stellar systems outside the Milky Way
\citep[e.g.][]{stetson14}. Together with the proper motion, the center-of-mass
velocity (or gamma velocity, $V_{\gamma}$) of this type of stars is thus a
fundamental parameter that should be derived with the highest possible accuracy.
With the Gaia satellite already measuring precise proper motions for thousands
of variable stars in the Milky Way and beyond, the missing ingredient in a full
determination of fundamental kinematic parameters for RR Lyrae stars is the
estimate of gamma velocity\footnote{Gaia is expected to measure radial
velocities with errors of 1--15~km~s$^{-1}$, depending on spectral type and
brightness, of stars with V$<$16~mag \citep{prusti16}, while proper motions will
have errors ranging from a few $\mu$as~yr$^{-1}$, for bright stars, down to a
few 100~$\mu$as~yr$^{-1}$, for stars down to V=20~mag.}. Indeed, the
uncertainties associated to the determination of gamma velocities of RR
Lyrae stars constitute a long-standing problem, which also affects the accuracy
of RR Lyraes as distance indicators \citep[e.g.][]{benedict11}. Several studies have been devoted to this
problem \citep{hawley85, jef07, for11, sessar12, nemec13}. Traditionally, the approach to derive the gamma velocity relies on a template radial velocity curve, that is shifted and scaled to match the observed radial velocity at a few different phases \citep{liu91}. This method becomes more accurate when observations covering several phases are available.
In this work, we apply a method which allows to estimate the gamma velocity
of RR Lyrae stars from just a few observations obtained at random phases. The method is
based on (i) investigations of the absorbtion line profile asymmetry that occurs
during the radial pulsations and (ii) determination of the absolute value of the
pulsation component using line profile bisectors, taking carefully into account
limb-darkening effects. We test our method on a sample of solar neighborhood RR
Lyraes that was investigated in a previous paper \citep[][hereinafter Paper
I]{pancino14} and on densely-spaced observations of RR Lyr \citep{fossati14}.

The paper is organized as follows: in Section 2 we briefly review the observations
and data reduction for our sample stars. In Section 3, we discuss the determination
of the radial velocity and gamma velocity when applying the `classic' method
based on radial velocity curve templates. Section 4 presents our method of bisectors for the
determination of the gamma and pulsation velocities of RR Lyraes, and
compares the results with those from the classical approach. Moreover, we tested our method on the high resolution observations of RR Lyr. 
Section 5 presents the results of the analysis; Section 6 closes the paper with the summary and conclusions. Appendix~A presents an extensive library of synthetic bisectors that can be used to apply our proposed method to spectra of variable stars.

\section{Observations and data reduction}
\label{sec-data}

The analysed sample consists of 20 RR Lyr stars and 3 Population II Cepheids.
The stars were observed as part of different programs with the
SARG \citep{sarg} echelle spectrograph at the Telescopio Nazionale Galileo (TNG,
La Palma, Spain) and with the UVES \citep{uves} spectrograph at ESO's Very Large
Telescope (VLT, Paranal, Chile). Furthermore, additional archival spectra were
retrieved from the ESO archive. The resolving power of the spectra obtained with
SARG is R = $\lambda/\delta \lambda \approx$ 30\,000 with an average
signal-to-noise ratio S/N $\approx$ 50-100 and spectral coverage from 4000 to
8500~\AA. The UVES spectra have a higher resolving power, R $\approx$ 47\,000, and
S/N $\approx$ 70-150 and cover the wavelength range from 4500 to 7500~\AA. The
observations were performed at random pulsation phases generally three times for
each star; however, for some stars we have more then 3 observations, or just one observation. Table
\ref{tab:info} presents some general information about the program stars.

A full description of the observations and reduction process, including the
determination of the fundamental parameters of the sample stars, is presented in
Paper I. Briefly, SARG spectra were reduced with standard IRAF\footnote{IRAF
(http://iraf.noao.edu/) is distributed by the National Optical Astronomical
Observatory, which is operated by the Association of Universities for Research in
Astronomy (AURA) under cooperative agreement with the National Science
Foundation.} tasks in the {\em echelle} package, including bias and flat-field
correction, spectral tracing, extraction, wavelength calibration, and continuum
normalization. The typical r.m.s. deviation of wavelength calibration lines
centroids from the two-dimensional fitted calibration polynomials was close to
0.03~\AA. Sky absorption lines (telluric bands of O$_{2}$ and H$_{2}$O) were
removed using the IRAF task {\em telluric} with the help of our own library of
observed spectra. UVES spectra were reduced with the UVES pipeline
\citep{uvespipe} as part of the service observations, and include similar
steps as the ones described for SARG.

The dataset from Paper~I was complemented by the densely spaced observations of
RR~Lyrae along the pulsation cycle described by \citet{fossati14}.

\section{Gamma velocity of pulsating stars}
\label{sec:classic}

From an observational point of view, the gamma velocity, $V_{\gamma}$,
can be described using the following equation:

\[
V_{\gamma}=v_{obs}+v_{\odot}-v_{puls} \qquad (1)
\]

\noindent where $v_{obs}$ is the observed velocity of the star along the line of
sight, $v_{\odot}$ is the heliocentric correction, and $v_{puls}$ is the
pulsation velocity of the radially pulsating star. We will see in the following
sections that the determination of $v_{puls}$ requires a treatment of
limb-darkening effects, generally included in the form of a {\em projection
factor}. The sum  $v_{obs}+v_{\odot}$  is the heliocentric radial
velocity of the star, $v_{rad}$, at any given phase in the pulsation cycle. In
other words, the determination of the gamma velocity of pulsating
stars such as RR Lyraes and Cepheids requires --- in principle --- the
determination of two quantities: (i) the observed radial velocity
(reported to the heliocentric reference) and (ii) the pulsation component at the
moment of the observation or, better, at the specific pulsation phase of the
observations (corrected for limb-darkening effects).

In the next subsections we apply the most widely used method for deriving the
gamma velocity to the sample described in Section~\ref{sec-data}. The
resulting measurements (reported in Table~\ref{tab:ind_ccf}) are then used as
reference values to test the method based on bisectors (Section~\ref{sec-new}).

\subsection{Measurement of $v_{obs}$ using cross-correlation}
\label{sec:CCF}

We first measured the radial velocity of our sample spectra using the
classical cross-correlation method, implemented in the IRAF {\em fxcor} task and
based on the standard \citet{fxcor} algorithm. We cross-correlated the observed spectra
with synthetic spectra generated with the latest modified version (15 March 2013)
of the STARSP-SynthV package, an LTE spectral synthesis code developed by \citet{T96}, which
uses ATLAS9 \citep{kuruch_atlas} grids of stellar atmosphere models. We adopted the
following typical atmospheric parameters of RR~Lyr stars for the synthetic spectrum
generation: $T_{\mathrm{eff}}$ = 6250~K, $\log g = 2.5$~dex,
$v_{mic}=2.0$~km~s$^{-1}$ -- as the typical values at quiescent phases along the RR lyrae stars pulsation cycle from Paper I. Then, we convolved the synthetic spectrum with a Gaussian to
reproduce the spectral resolution of the observed spectra (Section~\ref{sec-data}).
We performed a cross-correlation analysis for all spectra in the wavelength range
5100 - 5400~$\AA$, where several prominent, not blended, Ti and Fe lines are
present. We remark that the low metallicity of RR Lyrae stars leads to a limited line blending.

The resulting radial velocity measurements derived by the cross-correlation method ($v_{obs}^{Xcor}$) and their errors, $\delta
v_{obs}^{Xcor}$ as estimated by the position and full width at half maximum of the cross-correlation
peak, are reported in Table~\ref{tab:ind_ccf}, along with their heliocentric
corrections, $v_{\odot}$.

\begin{figure}
\resizebox{\hsize}{!}{\includegraphics[angle=0]{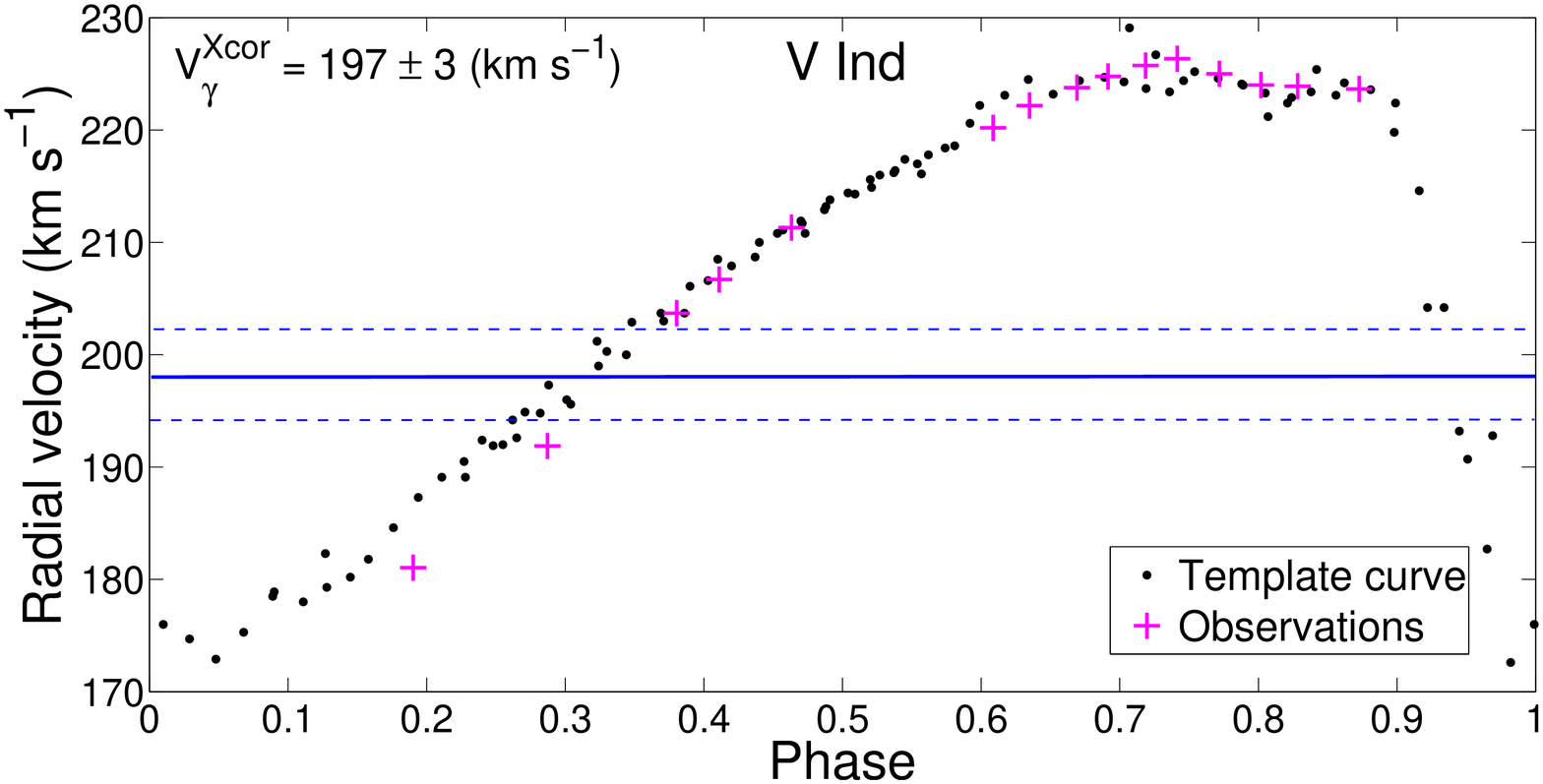}}
\resizebox{\hsize}{!}{\includegraphics[angle=0]{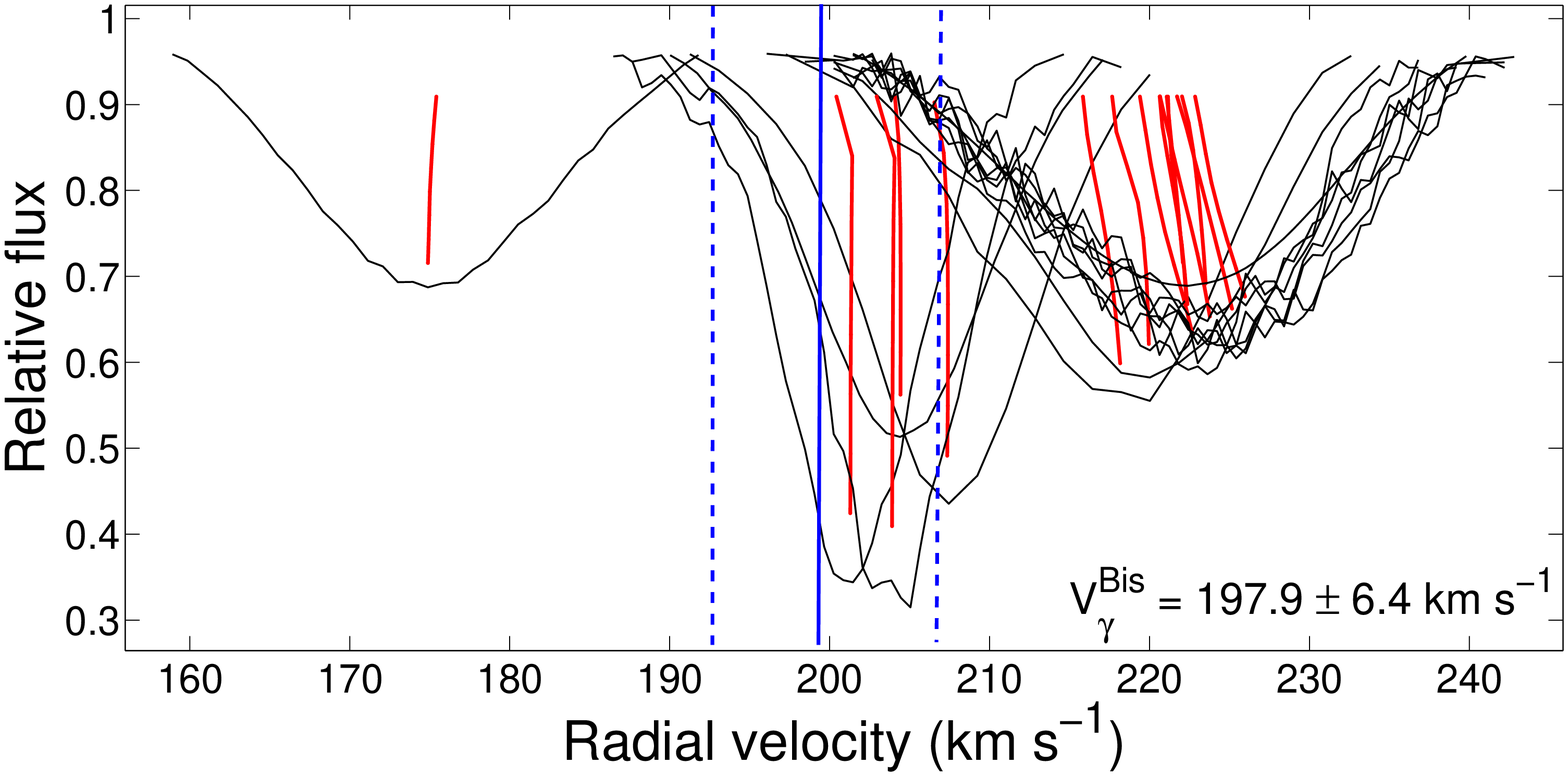}}
\resizebox{\hsize}{!}{\includegraphics[angle=0]{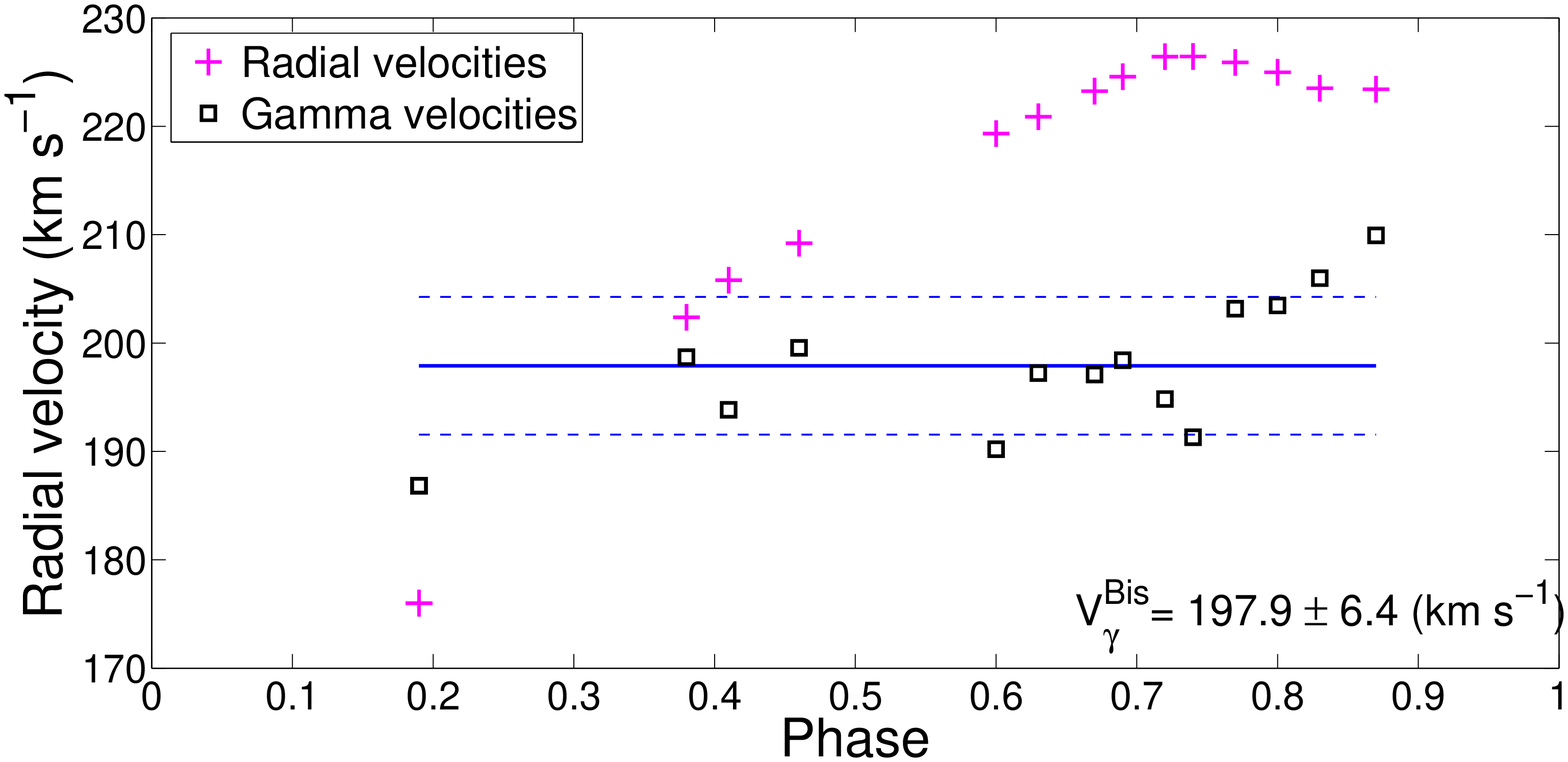}}
\caption[]{Examples of gamma velocity estimates for V~Ind, obtained with a
reference radial velocity curve \citep[][top panel, see also
Section~\ref{sec:classic}]{vind} and with the bisectors method (middle and bottom panel,
see Section~\ref{sec-new}), which has several spectra obtained at different phases.
In the middle panel we present the individual LSD profiles together with their bisectors, the bottom panel shows the observed radial velocities (pink crosses) and derived individual values of gamma velocity (black boxes). The resulting gamma velocity is marked by blue solid line in all panels, the dashed lines showing the errors associated to each determination.}
\label{fig:curve}
\end{figure}

\subsection{Measurement of $V_{\gamma}$ using velocity curve templates}

We used the classical radial velocity curve template approach to obtain the
final $V_{\gamma}^{Xcor}$, which is also reported in Table~\ref{tab:ind_ccf}. The
method relies on the use of an appropriate template radial velocity curve, that
is fitted to the observed radial velocity points, to obtain the systemic or
gamma velocity of the variable star.
In our sample we have RRab Lyrae (fundamental mode pulsators), and RRc Lyrae (first overtone pulsators). The former present asymmetric lightcurves with large radial velocity and magnitude ($\delta V\approx$ 0.5--2 mag) variations. The latter, instead, have relatively symmetric lightcurves with smaller magnitude and radial velocity variations. The physics of RRc type variability is still poorly understood \citep[see conclusions in][]{moskalik}, mainly because of the presence of several radial and non radial pulsation modes. Thus, in our analysis for RRc type stars we used as a reference the average radial velocity curves of well known RRc Lyraes with similar radial velocity amplitudes.
For a few of our RRab stars, we could rely on existing radial velocity curves
from the literature: V~Ind \citep{vind}; X~Ari \citep{Xari}; TU~Uma \citep{tu_uma}; U~Com \citep{beers00}.
For these stars the Baade-Wesselink (B-W) method was applied in the literature, thus the reliable radial velocity curves are exist for this sample. We noted that the typical amplitudes of these curves were all around 60 km~s$^{-1}$. For the RRc type variables in our sample, we used the velocity curves of DH~Peg \citep{dhpeg} and YZ~Cap \citep{swdra_yzcap}, both with  amplitudes of $\simeq$25~km~s$^{-1}$. For the three Cepheids, TW~Cap, V716~Oph, and UY~Eri we used the radial velocity curve of W~Vir \citep{wvir}.

\begin{table*}
\caption{Radial and gamma velocities of program stars, derived with the classical cross-correlation and radial velocity curves approach.}
\label{tab:ind_ccf}
\begin{tabular}{lcccrcrrrr}
\hline
Star & Inst.  & Exp &Phase &$v_{obs}^{Xcor}$ & $\delta v_{obs}^{Xcor}$ & $v_{\odot}$   & $v_{rad}^{Xcor}$ & $V_{\gamma}^{Xcor}$ & $\delta V_{\gamma}^{Xcor}$\\
      &      &      &      & (km~s$^{-1}$) &      (km~s$^{-1}$)       & (km~s$^{-1}$) & (km~s$^{-1}$)  & (km~s$^{-1}$) & (km~s$^{-1}$)\\
\hline
DR And    & SARG& 1& 0.63 &--119.67& 0.25 & 17.59  &--102.08 & --109.7 & 3.0 \\
          & SARG& 2& 0.69 &--116.33& 0.26 & 17.53  &--98.80  & --125.4 & 3.1 \\
          & SARG& 3& 0.31 &--118.91& 0.12 & 6.91   &--112.00 & --123.1 & 3.4 \\
X Ari     & APO & 1& 0.19 &--23.25 & 0.26 &--27.19 &--50.44  & --36.1 & 3.6 \\
TW Boo    & SARG& 1& 0.61 &--83.20 & 0.12 & 3.11   &--80.09 & --101.1 & 1.5\\
          & SARG& 2& 0.65 &--80.67 & 0.13 & 3.11   &--77.56 & --100.4& 1.6 \\
          & SARG& 3& 0.69 &--79.87 & 0.13 & 3.07   &--76.81 & --100.3& 1.4 \\
TW Cap    & UVES& 1& 0.54 &--46.60 & 0.13 &--15.68 &--62.28 & --72.3 & 5.0 \\
RX Cet    & SARG& 1& 0.51 &--75.79 & 0.22 & --4.34 &--80.13 & --93.7 & 3.1 \\
U Com     & SARG& 1& 0.06 &--13.46 &0.17  &--5.55  &--19.01  & --7.8 & 3.2 \\
          & SARG& 2& 0.14 &--8.42  & 0.14 &--5.58  &--14.00  & --5.6 & 3.2 \\
          & SARG& 3& 0.21 &--4.34  & 0.11 &--5.61  &--9.95   & --7.1 & 3.2 \\
RV CrB    & SARG& 1& 0.35 &--147.40& 0.19 & 13.70  &--133.70 & --138.3 & 3.6 \\
          & SARG& 2& 0.42 &--146.20& 0.28 & 13.66  &--132.54 & --139.8 & 3.7 \\
          & SARG& 3& 0.48 &--147.40& 0.70 & 13.62  &--133.78 & --142.4 & 3.7 \\
SW CVn    & SARG& 2& 0.21 &19.19   & 0.30 & --7.80 & 11.39 & 24.0 & 3.9  \\
          & SARG& 3& 0.28 &17.44   & 0.49 & --7.86 & 9.58  & 16.5 & 3.1  \\
UZ CVn    & SARG& 1& 0.04 &--39.35 & 0.18 &--8.70  &--47.44 & --19.6 & 4.1 \\
          & SARG& 2& 0.08 &--33.10 & 0.11 &--8.77  &--41.52 & --17.1 & 3.5\\
          & SARG& 3& 0.13 &--29.41 & 0.12 &--8.84  &--37.90 & --18.2 & 4.0 \\
AE Dra    & SARG& 1& 0.05 &--309.69& 0.18 & 6.03   &--303.66 & --275.7 & 1.7 \\
          & SARG& 2& 0.10 &--304.91& 0.17 & 5.99   &--298.92 & --275.3 & 1.7 \\
BK Eri    & UVES& 1& 0.04 & 95.33  & 0.69 &--27.12 & 68.20   & 98.6& 1.7 \\
          & UVES& 2& 0.20 & 109.29 & 0.16 &--27.03 & 82.26   & 97.2& 1.5 \\
          & UVES& 3& 0.14 & 102.80 & 0.21 &--27.04 & 75.76   & 95.8& 1.7 \\
          & UVES& 4& 0.51 & 137.82 & 0.08 &--26.83 & 110.99  & 96.3& 1.8 \\
UY Eri    & SARG& 1& 0.38 & 129.09 & 0.24 & 21.16  &150.25  & 151.9 &  5.0 \\
SZ Gem    & SARG& 1& 0.50 & 309.44 & 0.23 & 26.93  & 336.37 & 321.6& 3.0  \\
VX Her    & SARG& 1& 0.86 &--370.87& 0.13 & 18.47  &--351.96 & --340.1 & 5.3\\
          & SARG& 2& 0.05 &--391.85& 0.98& 18.43   &--373.33 & --402.4 & 4.8 \\
DH Hya    & UVES& 1& 0.79 & 389.03 & 0.18 &--24.17 & 364.86 & 336.3 & 4.5 \\
          & SARG& 2& 0.71 & 381.25 & 0.15 &--20.06 & 361.18 & 334.1 & 5.2 \\
          & SARG& 3& 0.79 & 385.19 & 0.26 &--20.17 & 365.02 & 334.3 & 4.3  \\
V Ind     & UVES& 1& 0.19 &175.00  & 0.22 &6.03    &181.03  & 194.9 & 4.6  \\
	      & UVES& 2& 0.29 &204.00  & 0.09 &--12.13 &191.87  & 196.1 & 3.3  \\
          & FEROS& 1& 0.46 &208.00  & 0.08 &3.31    &211.31 & 197.1 & 3.1 \\
          & FEROS& 2& 0.60 &217.00  & 0.11 &3.20    &220.20 & 198.3 & 3.0\\
          & FEROS& 3& 0.63 &219.00  & 0.11 &3.18    &222.18  & 198.0 & 3.1   \\
          & HARPS& 1& 0.67 &221.00  & 0.24 &2.77    &223.77  & 196.3 & 3.0 \\
          & HARPS& 2& 0.69 &222.00  & 0.24 &2.77    &224.77  & 197.6 & 3.0 \\
          & HARPS& 3& 0.72 &223.00  & 0.26 &2.74    &225.74  & 198.2 & 3.0 \\
          & HARPS& 4& 0.38 &201.00  & 0.17 &2.69    &203.69  & 198.9 & 3.2 \\
          & HARPS& 5& 0.41 &204.00  & 0.21 &2.69    &206.69  & 199.3 & 3.8 \\
          & HARPS& 6& 0.74 &224.00  & 0.28 &2.34    &226.34  & 198.0 & 4.9 \\
          & HARPS& 7& 0.77 &223.00  & 0.33 &2.00    &225.00  & 196.2 & 6.1 \\
          & HARPS& 8& 0.80 &222.00  & 0.33 &2.00    &224.00  & 194.4 & 9.8 \\
          & HARPS& 9& 0.83 & 222.00 & 0.27 &1.90    & 223.90 & 193.2 & 8.4 \\
          & HARPS&10& 0.87 & 221.78 & 0.21 &1.87    & 223.65 & 194.5 & 7.5  \\
SS Leo    & UVES & 1& 0.06 & 156.83 & 0.23 &--20.77 & 136.06 & 165.4 & 4.1 \\
          & UVES & 2& 0.07 & 155.53 & 0.25 &--20.79 & 134.74 & 162.2 & 3.9 \\
V716 Oph  & UVES& 1& 0.08 &--307.06& 0.11 &--28.25 &--335.31 & --310.8 & 5.0   \\
VW Scl    & UVES& 1& 0.71 & 86.85  & 0.12 &--25.91 & 67.02 & 40.3 & 3.2 \\
          & UVES& 2& 0.23 & 57.41  & 0.09 &--25.83 & 31.58 & 41.0 & 3.0 \\
          & UVES& 3& 0.32 & 65.86  & 0.06 &--25.59 & 40.27 & 41.2 & 3.1 \\
          & UVES& 4& 0.15 & 47.69  & 0.15 &--25.41 & 22.28 & 42.2 & 3.0 \\
          & UVES& 5& 0.05 & 37.76  & 0.37 &--25.63 & 12.13 & 37.7 & 3.2  \\
BK Tuc    & UVES& 1& 0.17 & 162.49 & 0.12 &--2.52  & 159.97  & 167.4 & 3.1 \\
          & UVES& 2& 0.26 & 162.25 & 0.15 &--2.57  & 159.68  & 174.4 & 3.1 \\
          & UVES& 3& 0.15 & 160.86 & 0.11 &--2.86  & 158.00  & 167.4 & 3.1 \\
TU UMa    & SARG& 1& 0.45 & 122.65 & 0.10 &--13.15 & 109.50  & 99.4  &  3.0 \\
RV Uma    & SARG& 1& 0.62 &--171.26& 0.16 &--5.61  &--176.87 &--200.1 &  3.9   \\
          & SARG& 2& 0.66 &--167.39& 0.14 &--5.64  &--173.03 & --198.8 & 3.7  \\
          & SARG& 3& 0.71 &--164.40& 0.15 &--5.67  &--170.07 & --196.2 & 5.9  \\
UV Vir    & UVES& 1& 0.87 &74.20   & 0.18 &--16.58 & 57.62   & 99.4  &  3.0  \\
\hline
\end{tabular}
\medskip
\end{table*}

For the remaining stars, we reconstructed synthetic radial velocity curves
following \citet{liu91}, who provided both a template radial velocity curve for
RRab stars and a correlation between the amplitudes of the light curves and
radial velocity curves, based on a large sample of RR~Lyrae observations. We
used the magnitude ranges reported in Table~\ref{tab:info} to obtain radial
velocity amplitudes with equation (1) by \citet{liu91}. We then divided the
obtained amplitude by an average projection factor of 1.3 (see also
Section~\ref{sec-dark}), and scaled the template curve appropriately. We then
adjusted the template curve to our measured radial velocities -- an example of
this procedure is shown in Figure~\ref{fig:curve} (top panel) for V~Ind, which
has several spectra at different phases. The final gamma velocity is
obtained as an average of the (scaled and shifted) radial velocity curve that
fits the observational points.

\begin{figure*}
\centering
\resizebox{\hsize}{!}{\includegraphics[angle=0]{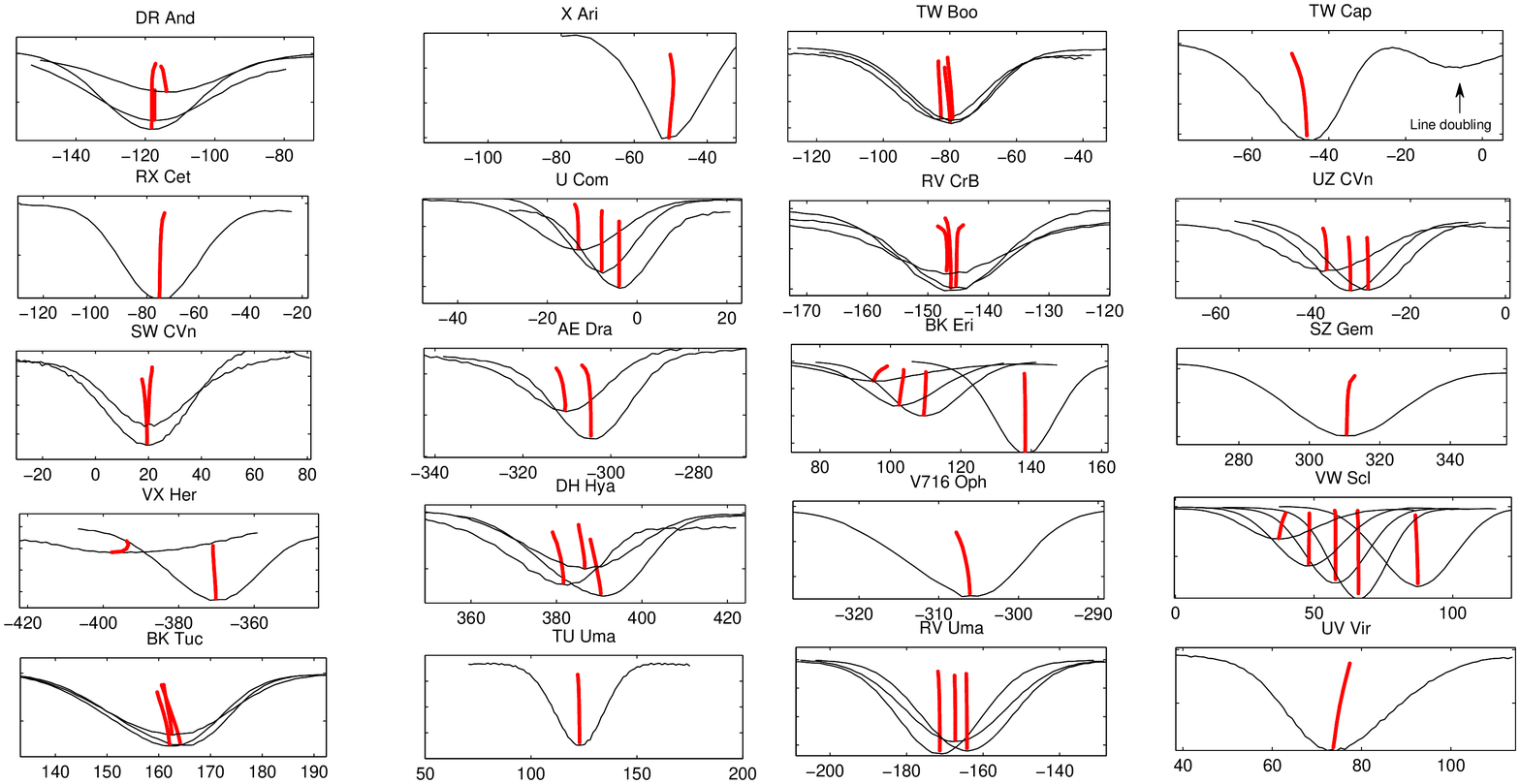}}
{\includegraphics[width=0.8\linewidth]{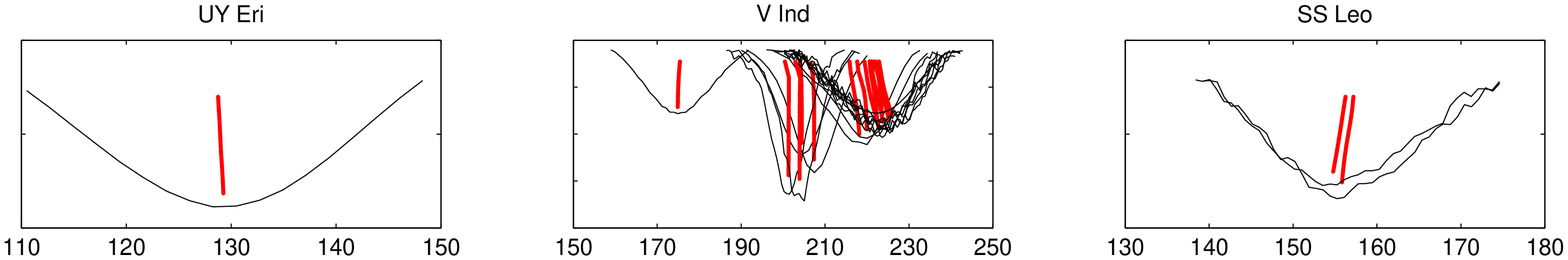}}
\caption[]{LSD profiles (black lines) and bisectors (red lines) for all sample spectra. The x-axis gives the velocity in km~s$^{-1}$.}
\label{fig:all_lsd}
\end{figure*}

The typical uncertainties connected to the use of this method depend on several factors. As discussed by
\citet[][see also \citealt{jef07}]{liu91}, the error in the normalization of the
radial velocity curve is about 3 km~s$^{-1}$, to which one has to add the individual radial velocity uncertainties (Table \ref{tab:ind_ccf}). Additionally, whenever an observed radial velocity curve was available in the
literature, we repeated the analysis with the observed curve and compared the
results. In each case, the results obtained with the two methods were comparable --- i.e., no systematic differences
were found --- so we used a weighted average of the results obtained with the
template and observed curve.
\section{Bisectors method}
\label{sec-new}

The classical method described above provides good results when a good phase
sampling is available, or, in case of few observations, when the phase of the observations is well known. In this section we discuss an alternative method for the determination of the gamma velocity,
based on the asymmetry of line profiles in pulsating stars, that vary along the
pulsation cycle and thus can also be used to infer the pulsational velocity. This method, hereafter referred to as {\em bisectors method}
for brevity, allows for an estimate of the gamma velocity of pulsating stars
with a sparse phase sampling, and it works even
with just one observation, albeit with a slightly larger uncertainty.
In the literature, this method was mentioned by several authors \citep[e.g.][]{hatzes96,gray10}. Other authors \citep[e.g.][]{sabey95, Nardetto2008, Nardetto2013} use the {\em Gaussian asymmetry coefficient} to estimate line asymmetries and perform the same task, however, the bisectors method allows for a more detailed study
of the line profile, i.e., to study line asymmetries and v$_{puls}$ variations at
different depths inside the stellar atmosphere. On the other hand, the bisectors method is in
general more sensitive to the spectra quality, i.e., S/N ratio, spectral resolution,
and line crowding and blending.

To derive average line profiles, we use the LSD method \citep[Least-Square
Deconvolution,][]{donati97}, that infers a very high S/N ratio line profile for
each spectrum from the profiles of many observed absorption lines, under the
assumption that the vast majority of lines have the same shape, and that
different line components add up linearly \citep{koch10}. Depending on the number of used lines,
the reconstructed LSD profile can have an extremely high S/N ratio, rarely
attainable with RR~Lyrae observations on single lines. The problem with RR Lyrae is that it is difficult to make long exposure observations without avoiding the line smearing effect, as a result its limit the S/N of obtained spectra. The original LSD method by \citet{donati97} was modified and extended to different applications by several
authors \citep[e.g.,][among others]{koch10,VR13,tka13}. We used the \citet{tka13}
implementation (see next section), and our workflow can be outlined as follows:

\begin{enumerate}
    \item{we compute the LSD profile of each observed spectrum; this also allows
    for an independent estimate of $v_{obs}$ (Section~\ref{sec-lsd});}
    \item{we compute a theoretical library of LSD profiles (Appendix~\ref{sec-lib}), predicting the line profile asymmetries at each pulsation phase, and compute the bisectors of each of the profiles; the
    theoretical bisectors library is made available in the electronic version of
    this paper and can be used to derive $v_{obs}$ and $V_{\gamma}$ for RR Lyrae observations obtained at unknown or poorly determined phases;}
    \item{we compute the bisector of the observed LSD line profile and compare
    that to the ones in our theoretical library, to determine which pulsation velocity
    corresponds to the observed asymmetry of the LSD bisector
    (Section~\ref{sec:puls}); in our computation, we implement a full
    description of limb-darkening effects (Section~\ref{sec-dark}).}
\end{enumerate}

Once the observed and pulsational components are known, the gamma velocity
can be trivially obtained from equation~(1).

\begin{table*}
\caption{Radial, pulsational and gamma velocities of program stars, derived with the LSD profile and the
bisector's method.}
\label{tab:ind_lsd}
\begin{tabular}{lcccrrrrrrr}
\hline
Star & Inst.  & Exp &Phase & $v_{obs}^{LSD}$ &  $\delta v_{obs}^{LSD}$ & $v_{rad}^{LSD}$-$V_{\gamma}^{Bis}$& $v_{puls}^{Bis}$ & $\delta v_{puls}^{Bis}$ & $V_{\gamma}^{Bis}$ & $\delta V_{\gamma}^{Bis}$ \\
      &      &      &      & ($km~s^{-1}$) & ($km~s^{-1}$) &  ($km~s^{-1}$)  &  ($km~s^{-1}$) 	   & ($km~s^{-1}$) & ($km~s^{-1}$) & ($km~s^{-1}$) \\
\hline
DR And    & SARG & 1& 0.63 & --119.8 & 0.5&    5.2 &    5 & 5.0 &--107.4 & 5.9 \\
          & SARG & 2& 0.69 & --115.0 & 0.9&  16.8 &   20  & 0.6 &--114.2 & 1.5 \\
          & SARG & 3& 0.31 & --118.6 & 0.5& --2.6 &  --5 & 1.0 &--109.0 & 3.3 \\
X Ari     & APO  & 1& 0.19 &  --51.1 & 1.3 &  --34.3 & --43 & 2.0  &--43.9 & 2.6 \\
TW Boo    & SARG & 1& 0.61 &  --83.0 & 1.7  & 8.1 &   10 & 5.0  &--88.0 & 5.4 \\
          & SARG & 2& 0.65 &  --80.3 & 1.2 & 11.4 &   13 & 1.5 &--88.6 & 2.1  \\
          & SARG & 3& 0.69 &  --79.2 & 0.2 & 14.3 &   17 & 1.9 &--90.4 &  2.2 \\
TW Cap    & UVES & 1& 0.54 &  --46.1 & 1.1 & 25.4 &   30 & 1.0  &--87.2 & 1.8 \\
RX Cet    & SARG & 1& 0.51 &  --75.5 & 0.5 & --11.8 & --17 & 2.1 &--68.0 & 2.4 \\
U Com     & SARG & 1& 0.87 &  --13.5 & 0.1  & 4.3 &    4 & 2.0 &--23.3 & 3.7 \\
          & SARG & 2& 0.95 &   --8.1 & 0.2  & 7.8 &   1 & 1.0  &--15.9 & 1.5 \\
          & SARG & 3& 0.02 &   --4.0 & 1.1  & 3.2 &    3 & 5.0 &--12.8 & 6.1 \\
RV CrB    & SARG & 1& 0.35 & --146.5 & 0.4 &   5.2 &    5 & 3.8 &--137.9 & 4.9 \\
          & SARG & 2& 0.42 & --145.3 & 0.8& --0.1 &  --2 & 6.3 &--131.5 &7.1 \\
          & SARG & 3& 0.48 & --147.3 & 0.2 & --3.1 &  --6 & 2.5 &--130.5 & 2.7 \\
SW CVn    & SARG & 2& 0.21 &    17.8 & 2.0 & --0.1 &  --2 & 3.3 &  10.1 & 5.4 \\
          & SARG & 3& 0.28 &    17.7 & 1.0& --7.7 & --12 & 2.9  & 17.6 & 3.2\\
UZ CVn    & SARG & 1& 0.04 &  --38.1 & 0.2 & --0.1 &  --2 & 5.0 &--46.0 & 5.9 \\
          & SARG & 2& 0.08 &  --32.9 & 0.2 & 14.2 &   17 & 0.5 &--55.9 & 1.2 \\
          & SARG & 3& 0.13 &  --29.0 & 0.1 & 7.9 &   10 & 0.5 &--45.7 & 1.2 \\
AE Dra    & SARG & 1& 0.05 & --309.3 & 0.6 & 11.3 &   13 & 1.0 &--314.5 & 1.6 \\
          & SARG & 2& 0.10 & --304.6 & 0.4 &  9.7 &   11 & 5.0 &--308.3 & 5.1 \\
BK Eri    & UVES & 1& 0.04 &    95.0 & 0.1 &--14.3 & --17 & 1.5  &  82.2 & 1.9 \\
          & UVES & 2& 0.20 &   109.2 & 0.1 &--11.8 & --16 & 0.5 &  93.9 & 1.2 \\
          & UVES & 3& 0.14 &   102.3 & 0.7 &--13.3 & --19 & 1.0 &  88.6 & 1.6 \\
          & UVES & 4& 0.51 &   138.1 & 0.2 &   2.7 &    1 & 3.0 & 108.6 & 4.3 \\
UY Eri    & SARG & 1& 0.38 &   129.3 & 1.1 &   4.4 &    4 & 4.5 &  146.1 & 5.7 \\
SZ Gem    & SARG & 1& 0.50 &   310.2 & 0.2 & --8.9 & --13 & 3.9 &  346.0 & 4.1 \\
VX Her    & SARG & 1& 0.86 & --370.7 & 0.8 &   7.1 &    8 & 5.0 &--359.3 & 5.2 \\
          & SARG & 2& 0.05 & --397.3 & 0.9 & --34.0 & --43 & 2.0 &--344.9 & 2.4 \\
DH Hya    & UVES & 1& 0.79 &   390.1 & 1.2 & 27.9 &   33 & 1.5  &338.1 & 2.1 \\
          & SARG & 2& 0.71 &   381.8 & 0.5 & 24.2 &   29 & 5.0  &337.5 & 5.1 \\
          & SARG & 3& 0.79 &   385.8 & 1.1 &--2.7 &  --5 &  8.0 &368.5 & 8.7 \\
V Ind     & UVES & 1& 0.19 &   174.9 & 0.1& --11.8 & --17 & 0.6&  192.7 & 1.3 \\
          & UVES & 2& 0.29 &   204.4 & 0.1 &  --1.9 &  --4 & 3.5 &  194.2 & 4.7 \\
          & FEROS& 1& 0.46 &   207.5 & 0.2 &   5.2 &    5  & 3.1 & 205.6 & 4.5 \\
          & FEROS& 2& 0.60 &   218.4 & 1.6  &  31.3 &   38 & 0.5 & 190.3 & 2.0 \\
          & FEROS& 3& 0.63 &   220.0 & 0.1 &  15.8 &   19 & 5.0 &  207.4 & 7.1 \\
          & HARPS& 1& 0.67 &   223.1 & 0.6 &  36.1 &   43 & 0.5 &  189.7 & 1.3 \\
          & HARPS& 2& 0.69 &   224.1 & 1.3 &  36.1 &   43 & 0.5 &  190.8 & 1.8  \\
          & HARPS& 3& 0.72 &   225.6 & 0.7 &  14.3 &   17 & 1.0 &  208.4 & 1.9 \\
          & HARPS& 4& 0.38 &   201.3 & 0.2 &--10.7 & --16 & 3.5 &  214.7 & 3.6\\
          & HARPS& 5& 0.41 &   203.9 & 1.2 & --7.2 & --11 & 1.9 &  213.9 & 2.4 \\
          & HARPS& 6& 0.74 &   226.5 & 1.0 &  36.1 &   43 &0.5  &  192.7 & 1.5\\
          & HARPS& 7& 0.77 &   223.7 & 1.3 &  15.7 &   19 & 0.5 &  210.0 & 1.7 \\
          & HARPS& 8& 0.80 &   222.5 & 1.2 &  19.5 &   24 & 1.0 & 205.0 & 1.9 \\
          & HARPS& 9& 0.83 &   222.5 & 0.5 &  22.3 &   27 & 3.7 &  202.1 & 3.8 \\
          & HARPS&10& 0.87 &   222.3 & 0.2 &  18.4 &   22 & 2.1 &  205.8 & 2.4 \\
SS Leo    & UVES & 1& 0.06 &  155.7  & 0.5 & --23.2 & --32 & 0.5  & 158.1& 1.1 \\
          & UVES & 2& 0.07 &  154.6  & 1.1 & --23.1 & --31 & 2.6  & 156.8& 3.0 \\
V716 Oph  & UVES & 1& 0.08 & --305.8 & 1.2 & 36.0  &   43 & 0.5 &--370.1 &1.7  \\
VW Scl    & UVES & 1& 0.71 &    87.2 & 1.5 &   5.2 &    4 & 0.5 &   56.1 & 3.8 \\
          & UVES & 2& 0.23 &    57.4 & 0.7 & --0.7 &  --1 & 2.5 &   32.3 & 4.0 \\
          & UVES & 3& 0.32 &    65.8 & 0.4 &   1.2 &    0 & 5.0 &   39.1 & 5.8 \\
          & UVES & 4& 0.15 &    47.6 & 0.3 & --7.3 &   11 & 3.5 &   29.4 & 3.6 \\
          & UVES & 5& 0.05 &    36.9 & 0.5 &--13.1 & --17 & 1.2  &   24.4 & 1.7 \\
BK Tuc    & UVES & 1& 0.17 &   164.3 & 0.2 &  33.8 &   40 & 1.7  &  128.0 & 2.0 \\
          & UVES & 2& 0.26 &   162.7 & 0.8 &  19.1 &   23 & 4.3 & 141.1 & 4.5 \\
          & UVES & 3& 0.15 &   162.1 & 1.0 &  25.6 &   31 & 2.2 &  133.7 & 2.6 \\
TU UMa    & SARG & 1& 0.45 &   122.9 & 0.6 &   3.2 &    3 & 2.5 &  106.5 & 4.0 \\
RV Uma    & SARG & 1& 0.62 & --171.6 & 0.1 &   2.7 &    1 & 1.5 &--179.9 & 3.4 \\
          & SARG & 2& 0.66 & --167.5 & 2.0 &   1.2 &    0 & 5.0 &--174.3 & 6.5 \\
          & SARG & 3& 0.71 & --164.2 & 0.4 &   1.2 &    0 & 5.0 &--171.1 & 5.8 \\
UV Vir    & UVES & 1& 0.87 &    73.4 & 0.2 & --23.4 & --32 & 2.0 &  80.8 & 2.3 \\
\hline
\end{tabular}
\end{table*}

\subsection{Measurement of $v_{obs}$ using LSD profiles}
\label{sec-lsd}

The LSD method for computing the line profile requires a list of spectral lines,
specified by their central peak position and depth, or central intensity, which
is generally referred to as a {\em line mask}. The whole spectrum is thus
modeled as the convolution of an ideal line profile (assumed identical for all
lines) with the actual pattern of lines from the mask. The LSD line profile is
obtained by deconvolution from the line pattern, given an observed spectrum.
Because the whole computation takes place in velocity space, the actual
LSD profile carries the $v_{obs}$ information, in our particular case determined as the center-of-mass of the profile
\citep{hareter08, koch10}.

We built our line mask from the same line list that was used for our abundance
analysis in Paper~I, and there described in detail. Briefly, we selected all
isolated lines that were well measured (errors below 20\%) in at least three of
the available spectra. For the line measurements we used DAOSPEC
\citep{daospec}. After the abundance analysis, performed with GALA \citep{gala},
all lines that gave systematically discrepant abundances, or that were
systematically rejected because of their strengths or relative errors, were
purged from the list. The final line list consisted of 352 isolated lines
belonging to 9 different chemical species (see Paper~I).

\begin{figure}
\resizebox{\hsize}{!}{\includegraphics[angle=0]{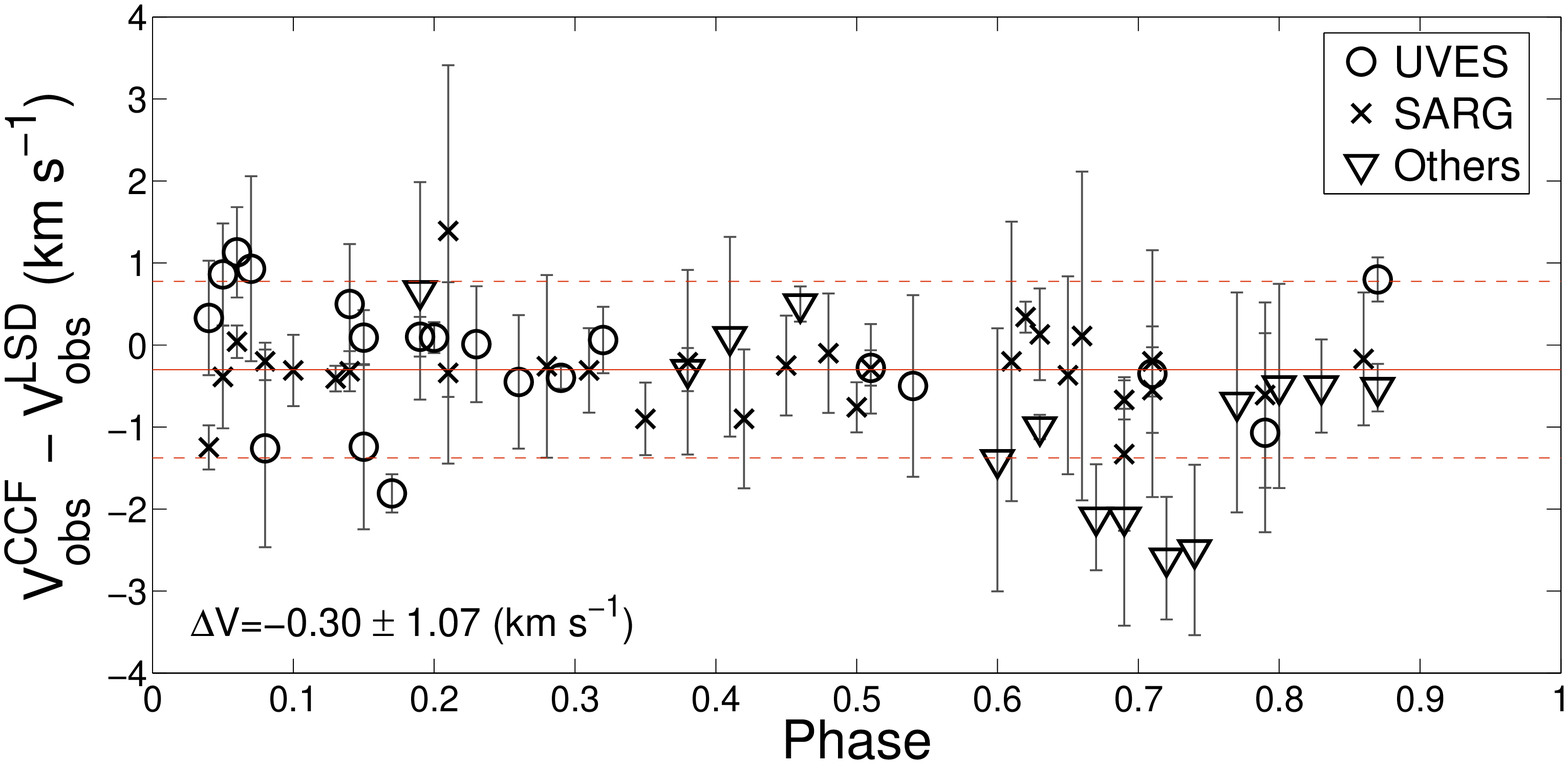}}
\caption[]{Differences between radial velocity estimates derived with the
cross-correlation method, $v_{obs}^{Xcor}$, and with the LSD profile method,
$v_{obs}^{LSD}$. The observations from the different spectrographs are labeled
respectively.}
\label{fig:rv_obs}
\end{figure}

We used the LSD code described by \citet[][]{tka13}, which is a generalization of
the original method by \citet{donati97}. We applied the method in the wavelength
range 5100--5400~$\AA$\footnote{ It is worth emphasizing that the wavelength
range 5100--5400 $\AA$ was used for both the cross-correlation analysis described
in Section~\ref{sec:CCF} and the LSD analysis, for consistency.}, that contains 155 of the above mentioned lines from Paper~I.

Figure~\ref{fig:all_lsd} displays the results of the computation. The resulting
$v_{obs}^{LSD}$ are reported in Table~\ref{tab:ind_lsd}, along with their errors,
$\delta v_{obs}^{LSD}$. To evaluate the final errors, we evaluated different
possible error sources and we summed them in quadrature. A first uncertainty is the difference between the minimum of the LSD
profile and its center-of-mass, typically in the range
0.5--1~km~s$^{-1}$. Since the line selection can have an impact on the line profiles, we repeated our measurements with three different line lists --- using lines with a depth lower than 65\%, 75\%, and 85\% of the continuum
level --- and used the dispersion in the resulting radial velocity as the
uncertainty caused by the line selection procedure, obtaining
values typically around 0.5~km~s$^{-1}$. We note that V~Ind has more observations
along the pulsation cycle than other stars in the sample, which makes it the
perfect test case for our method. The LSD profiles for V~Ind are displayed in the
lower panel of Figure~\ref{fig:curve}, where the variation of the line shape with phase is
clearly visible in both the line shape asymmetry and the bisectors slope.

Finally, we compared the $v_{obs}$ obtained with the classical cross-correlation
method with those obtained with the LSD profile method
(Figure~\ref{fig:rv_obs}), and found an extremely good agreement, with an average
offset of 0.3~km~s$^{-1}$ and a spread generally below 1~km~s$^{-1}$, which is
compatible with the individual error estimates of the two methods.

\begin{figure}
\resizebox{\hsize}{!}{\includegraphics[angle=0]{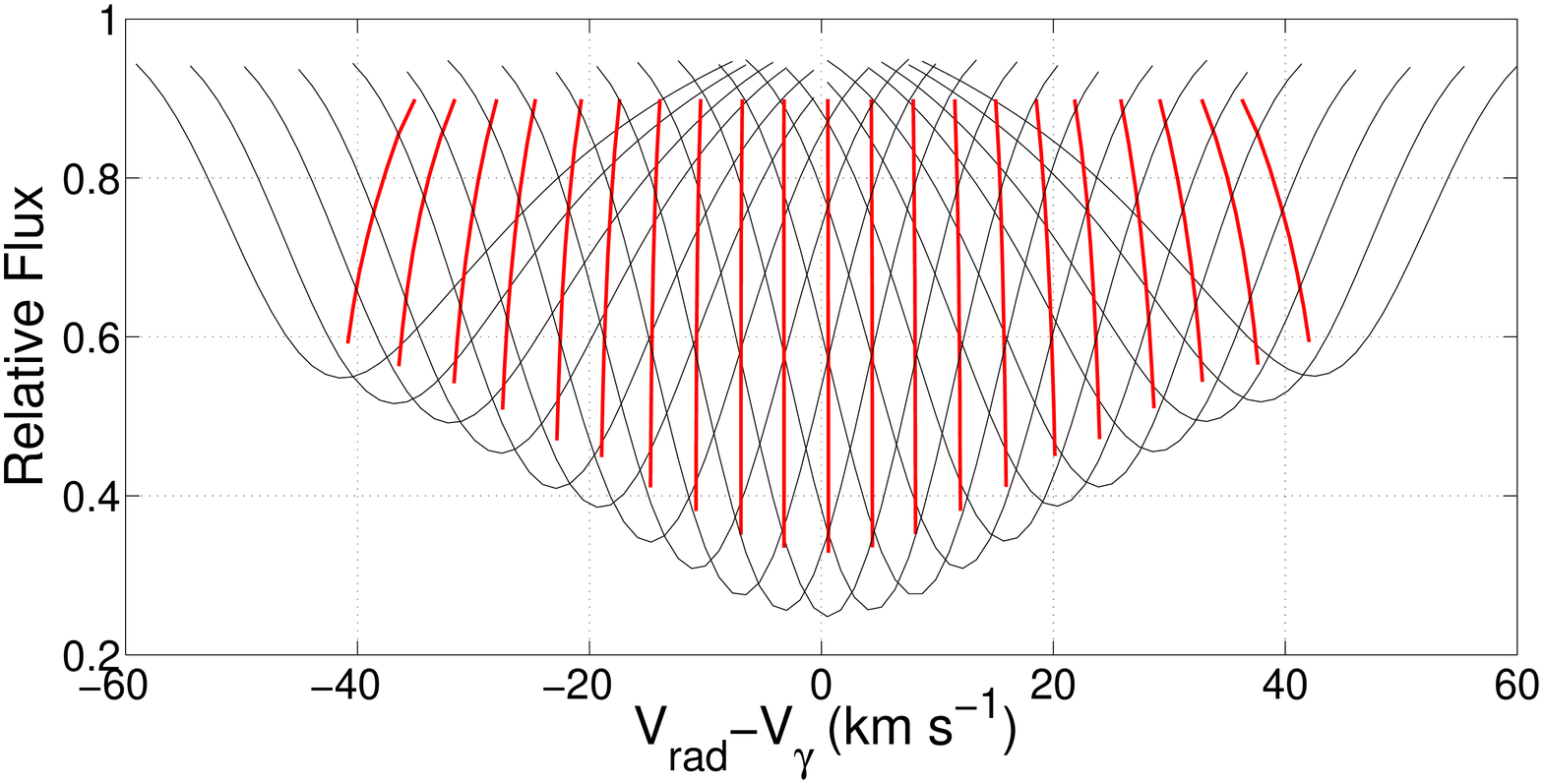}}
\caption[]{Example of LSD profiles and bisectors computed from model spectra of
RR~Lyrae \citep[that corresponds to $\phi$=0.213, or spectrum number 087
in][]{fossati14} with pulsation velocities from $-50$ to $+50$~km~s$^{-1}$ with a step $5$~km~s$^{-1}$.}
\label{fig:template_lsd}
\end{figure}

The star TW Cap in our sample deserves particular attention. This star is a well
known peculiar Type II Cepheid \citep{maas07}. The LSD profile shows a double
minimum (see Figure~\ref{fig:all_lsd}), hinting at the line doubling phenomenon.
Moreover, a double emission peak in $H_{\beta}$ is clearly visible in our
spectrum. \cite{twcap1958} was the first to report about double lines in the
spectra of TW~Cap. These features could be explained by the existence of
supersonic shock waves in the atmosphere at our observation phase ($\phi =
0.54$). The same phenomenon is observed during shock phases in the spectra of
RR~Lyr \citep{chadid08}.

\begin{figure}
\resizebox{\hsize}{!}{\includegraphics[angle=0]{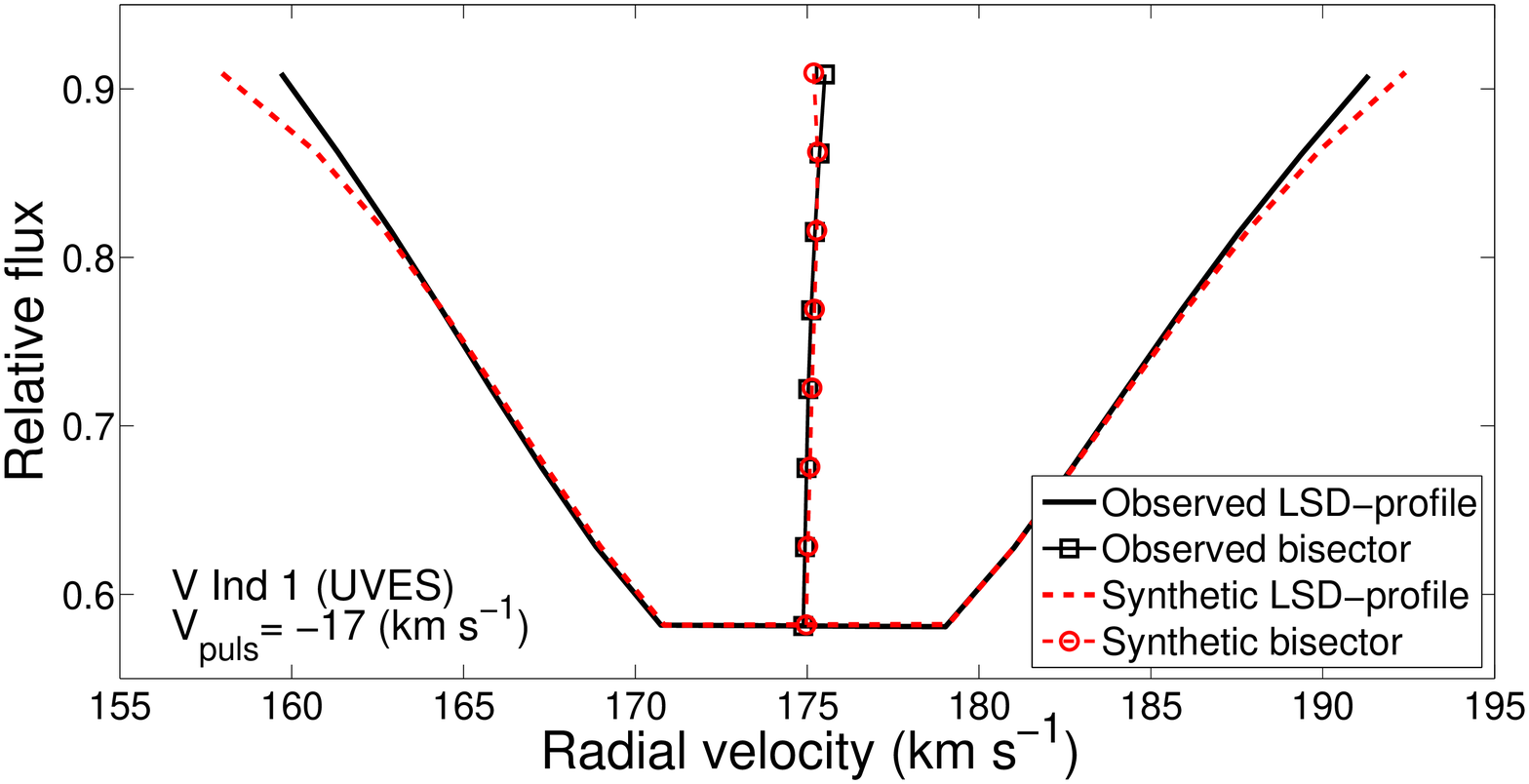}}
\resizebox{\hsize}{!}{\includegraphics[angle=0]{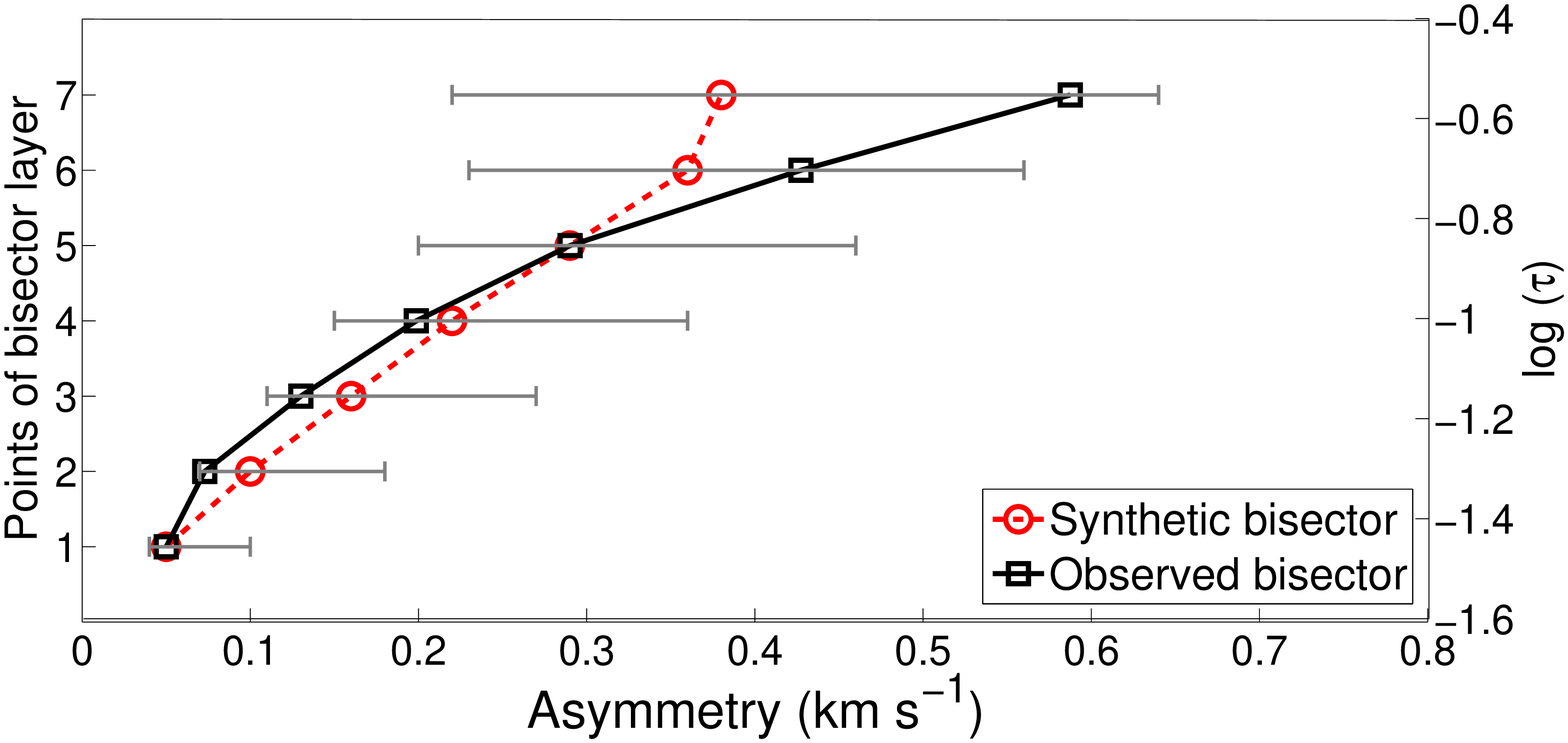}}
\caption[]{Comparison of the observed (black solid) and best fitting theoretical
(red dashed) LSD profiles (top panel) and bisectors (bottom panel) for our first
V~Ind spectrum ($\phi$=0.19, see Table~\ref{tab:ind_lsd}). The best fit
corresponds to $v_{puls}$=--17$\pm$1~km~s$^{-1}$. The grey error
bars in the bottom panel correspond to the estimated errors on the method for
determining $v_{puls}$, i.e., 1~km~s$^{-1}$. The optical depth
scale on the right ordinates axis was computed from synthetic spectra built with
the atmospheric parameters derived in Paper~I: $T_{\mathrm{eff}}$=7000~K,
$\log~g=2.3$~dex, $v_{\mathrm{mic}}=1.6$~km~s$^{-1}$, and [Fe/H]=$-$1.3~dex. The
divergence between the observed and theoretical bisectors on layer 7 are caused
by the lower S/N ratio of the LSD profile in this layer (see text for details). While the
model bisector does not follow exactly the shape of the observed ones, the two agree reasonably well within the uncertainties.}
\label{fig:lsd_vind}
\end{figure}

\subsection{Measurement of $v_{puls}$ using line bisectors}
\label{sec:puls}

For a quantitative characterization of the LSD profile asymmetry, we used the bisector technique, which is
a widely used diagnostic of line asymmetries \citep[e.g.][]{gray10}. For each LSD
profile, we computed the bisector of eight equally spaced depth layers, starting
from a level of relative flux equal to 0.91 and reaching to the bottom of the
line profile, at a flux level varying from 0.46 to 0.77, depending on the
spectrum. The upper layer of the LSD profile, however, was found to be relatively
uncertain, partly because we restricted ourselves to a small
wavelength region and partly because of the paucity of lines in some of the
analyzed spectra, caused by the low metallicity. Therefore, in the
following, only the lowest seven layers will be employed (with relative flux
below 0.85). The observed bisectors are shown in Figure~\ref{fig:all_lsd}
together with the computed LSD profiles.

To derive the pulsational velocity, $v_{puls}$, we compared the observed
bisectors with a library of theoretical bisectors specifically computed for the
program stars. The library is described in details in Appendix~\ref{sec-lib} and
is publicly available in the electronic version of this paper and in CDS. One
grid of theoretical LSD profiles and bisectors is shown on
Figure~\ref{fig:template_lsd}, as an example. To find the best-matching
theoretical bisector for each observed bisector, we minimized the quadratic
difference between the two $$\sqrt{\sum_{n=1}^{n=7} (Bis_{n}^{Obs} -
Bis_{n}^{Templ})^{2}}$$ \noindent where 7 is the number of bisector points we
take into account in the analysis (see above), and $Bis_{n}^{Obs}$ and
$Bis_{n}^{Templ}$ are the values of the bisector evaluated at each layer of the
observed and theoretical LSD profiles, respectively. In those cases where
the exact phase of the observation is unknown, one can proceed in two steps.
First, a blind comparison with the entire library is performed, providing an
indication of the phase, and then a second iteration with the appropriate
bisectors for that phase can be run.

Figure~\ref{fig:lsd_vind} illustrates, as an example, the best fit obtained with this technique for our first exposure of
V~Ind (see Table~\ref{tab:ind_lsd}).

\subsection{A test on RR~Lyr} \label{rr_lyr_sect}

In order to test the method, we applied it to the prototype of the RR~Lyrae
class: RR~Lyr itself. We used the 41 high-resolution (R=60\,000) and high
signal-to-noise (100--300) spectra of RR~Lyr taken along the whole pulsation
cycle analysed by \citet{kolen10,fossati14}. In the following papers authors determined the phase dependent atmospheric parameters of RR Lyr based on which we compute the synthetic spectra and theoretical bisectors for each of 41 observations of RR Lyr. Thus, the number and IDs of each spectrum of RR Lyr from \citet{fossati14} are the same as the number of theoretical stellar atmosphere models which we used for constructing the grid of synthetic bisectors (see Appendix~\ref{sec-lib}). The good phase sampling allows us to test the method thoroughly along the whole pulsation cycle. In the table \ref{tab:rrlyr} the first three column present basic information about the set of RR Lyr spectra from \citet{kolen10,fossati14} which we used for computing synthetic spectra and constructing our grids of bisectors.

\begin{table*}
\caption{Radial velocity measurements for the 41 spectra of RR~Lyr from \citet{kolen10}.}
\centering
\label{tab:rrlyr}
\begin{tabular}{lcccrrrrrrr}
\hline
 Number & Spectrum ID* & Phase & $T_{eff}$** & v$_{obs}^{LSD}$ & $\delta$v$_{obs}^{LSD}$ & v$_{puls}$ & v$_{puls}$/(p-factor) &  V$_{\gamma}^{Bis}$ & $\delta$V$_{\gamma}^{Bis}$ & p-factor  \\
  & & & &  (km s$^{-1}$)  &  (km~s$^{-1}$)  & (km~s$^{-1}$) & (km~s$^{-1}$) & (km s$^{-1}$)  &  (km~s$^{-1}$)  &  \\
\hline
1  & 87  & 0.173 &6325 $\pm$ 50  &--93.05 & 0.87 & --11 & --7.69 & --85.36 & 2.67 & 1.43 \\
2  & 88  & 0.207 &6275 $\pm$ 50  &--89.56 & 0.65 & --11 & --7.74 & --81.83 & 6.86 & 1.42 \\
3  & 89  & 0.229 &6225 $\pm$ 50  &--87.27 & 0.58 & --11 & --7.74 & --79.53 & 7.76 & 1.42 \\
4  & 91  & 0.260 &6175 $\pm$ 100 &--83.20 & 0.59 & --11 & --7.74 & --75.46 & 6.85 & 1.42 \\
5  & 120 & 0.846 &6125 $\pm$ 100 &--59.12 & 1.50 & 19 & 15.11 & --74.23 & 1.50 & 1.26 \\
6  & 121 & 0.868 &6375 $\pm$ 100 &--67.33 & 0.95 & 3 & 3.02 & --70.35 & 0.88 & 0.99 \\
7  & 122 & 0.890 &6725 $\pm$ 100 &--87.20 & 0.06 & --30 & --22.56 & --64.64 & 0.69 & 1.33 \\
8  & 124 & 0.922 &7050 $\pm$ 100 &--108.97 & 0.67 & --44 & --34.78 & --74.19 & 2.31 & 1.27 \\
9  & 125 & 0.943 &7125 $\pm$ 100 &--114.35 & 0.32 & --50 & --40.54 & --73.81 & 1.70 & 1.23 \\
10 & 126 & 0.967 &7050 $\pm$ 100 &--114.00 & 0.41 & --43 & --33.67 & --80.33 & 1.31 & 1.28 \\
11 & 158 & 0.604 &6000 $\pm$ 50  &--57.63 & 0.56 & 20 & 15.95 & --73.58 & 0.45 & 1.25 \\
12 & 160 & 0.647 &6050 $\pm$ 50  &--57.26 & 0.57 & 21 & 16.74 & --74.00 & 0.50 & 1.25 \\
13 & 161 & 0.669 &6050 $\pm$ 50  &--56.92 & 0.85 & 24 & 19.11 & --76.02 & 0.78 & 1.26 \\
14 & 163 & 0.698 &6025 $\pm$ 50  &--56.67 & 1.01 & 27 & 21.51 & --78.19 & 1.68 & 1.26 \\
15 & 164 & 0.720 &6050 $\pm$ 50  &--56.78 & 0.66 & 26 & 20.55 & --77.33 & 0.70 & 1.26 \\
16 & 165 & 0.741 &6025 $\pm$ 50  &--56.30 & 0.52 & 27 & 21.51 & --77.81 & 1.44 & 1.26 \\
17 & 166 & 0.763 &6000 $\pm$ 50  &--55.75 & 0.61 & 26 & 20.57 & --76.32 & 0.56 & 1.26 \\
18 & 168 & 0.792 &6025 $\pm$ 50  &--55.04 & 0.73 & 26 & 20.55 & --75.59 & 0.24 & 1.26 \\
19 & 169 & 0.814 &6025 $\pm$ 50  &--55.09 & 1.01 & 27 & 21.57 & --76.66 & 1.23 & 1.25 \\
20 & 170 & 0.834 &6050 $\pm$ 50  &--56.82 & 1.55 & 24 & 19.09 & --75.90 & 1.56 & 1.26 \\
21 & 171 & 0.856 &6275 $\pm$ 75  &--62.12 & 1.53 & 20 & 15.99 & --78.11 & 1.34 & 1.25 \\
22 & 174 & 0.905 &6925 $\pm$ 100 &--78.32 & 0.42 & --25 & --18.90 & --59.42 & 3.29 & 1.32 \\
23 & 175 & 0.928 &7125 $\pm$ 75  &--110.98 & 0.89 & --47 & --38.34 & --72.64 & 0.37 & 1.23 \\
24 & 176 & 0.948 &7125 $\pm$ 75  &--113.85 & 0.06 & --49 & --39.89 & --73.97 & 0.88 & 1.23 \\
25 & 204 & 0.349 &6050 $\pm$ 50  &--73.42 & 1.05 & 11 & 8.85 & --82.27 & 1.61 & 1.24 \\
26 & 205 & 0.372 &6000 $\pm$ 50  &--71.09 & 1.13 & 14 & 11.04 & --82.13 & 9.44 & 1.27 \\
27 & 206 & 0.394 &5950 $\pm$ 75  &--69.65 & 1.20 & 17 & 13.47 & --83.12 & 0.88 & 1.26 \\
28 & 207 & 0.416 &5950 $\pm$ 150 &--66.77 & 1.69 & 27 & 21.51 & --88.27 & 1.68 & 1.26 \\
29 & 209 & 0.452 &5975 $\pm$ 75  &--63.70 & 1.43 & 26 & 20.57 & --84.27 & 1.58 & 1.26 \\
30 & 210 & 0.475 &5950 $\pm$ 75  &--62.56 & 1.29 & 20 & 15.93 & --78.49 & 2.37 & 1.26 \\
31 & 251 & 0.098 &6525 $\pm$ 50  &--98.97 & 0.85 & --21 & --15.61 & --83.36 & 1.95 & 1.35 \\
32 & 252 & 0.120 &6450 $\pm$ 50  &--96.71 & 0.82 & --17 & --12.31 & --84.40 & 2.10 & 1.38 \\
33 & 253 & 0.141 &6400 $\pm$ 75  &--94.90 & 0.78 & --15 & --10.82 & --84.08 & 1.52 & 1.39 \\
34 & 255 & 0.173 &6325 $\pm$ 75  &--90.86 & 0.56 & --19 & --13.90 & --76.95 & 2.17 & 1.37 \\
35 & 256 & 0.203 &6250 $\pm$ 50  &--87.01 & 0.49 & --17 & --12.30 & --74.71 & 1.65 & 1.38 \\
36 & 257 & 0.226 &6200 $\pm$ 75  &--85.45 & 0.74 & --19 & --13.84 & --71.60 & 0.93 & 1.37 \\
37 & 258 & 0.249 &6175 $\pm$ 50  &--82.63 & 0.44 & --17 & --12.18 & --70.46 & 1.72 & 1.40 \\
38 & 260 & 0.278 &6100 $\pm$ 50  &--79.65 & 0.69 & --7 & --4.41 & --75.24 & 4.33 & 1.59 \\
39 & 261 & 0.303 &6050 $\pm$ 75  &--76.65 & 0.91 & --9 & --6.11 & --70.54 & 1.90 & 1.47 \\
40 & 262 & 0.327 &6025 $\pm$ 50  &--74.53 & 1.13 & 15 & 12.01 & --86.54 & 2.30 & 1.25 \\
41 & 263 & 0.349 &6025 $\pm$ 50  &--72.57 & 1.23 & 18 & 14.17 & --86.74 & 2.24 & 1.27 \\
\hline
\end{tabular}
\medskip \begin{flushleft} \emph{Notes.} * The IDs of spectra are according to the \citet{kolen10} and \citet{fossati14}. ** The effective temperature estimations are from \citet{fossati14}.
\end{flushleft}

\end{table*}

\begin{figure*}
\includegraphics[scale=0.3,angle=0]{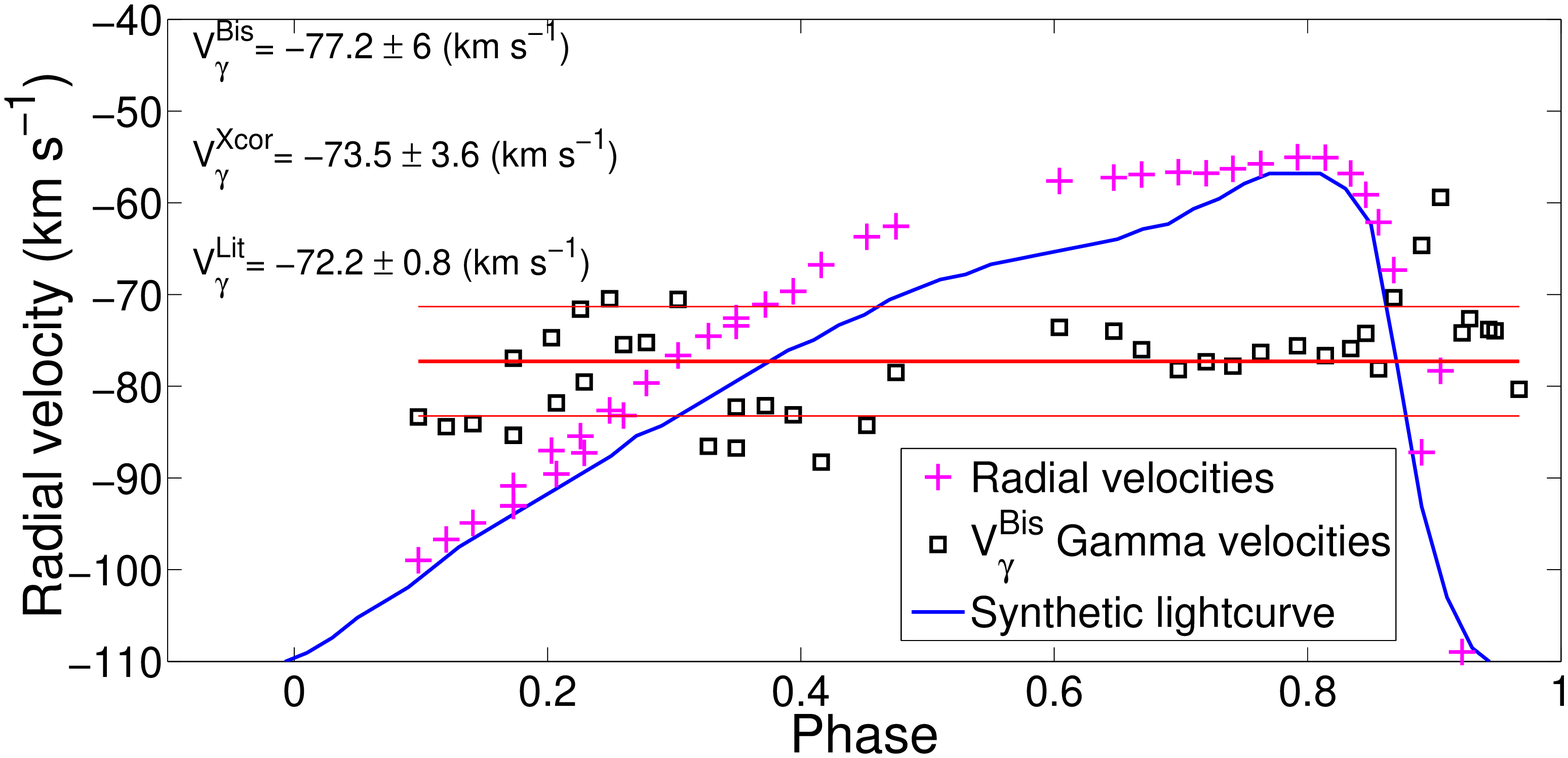}
\includegraphics[scale=0.3,angle=0]{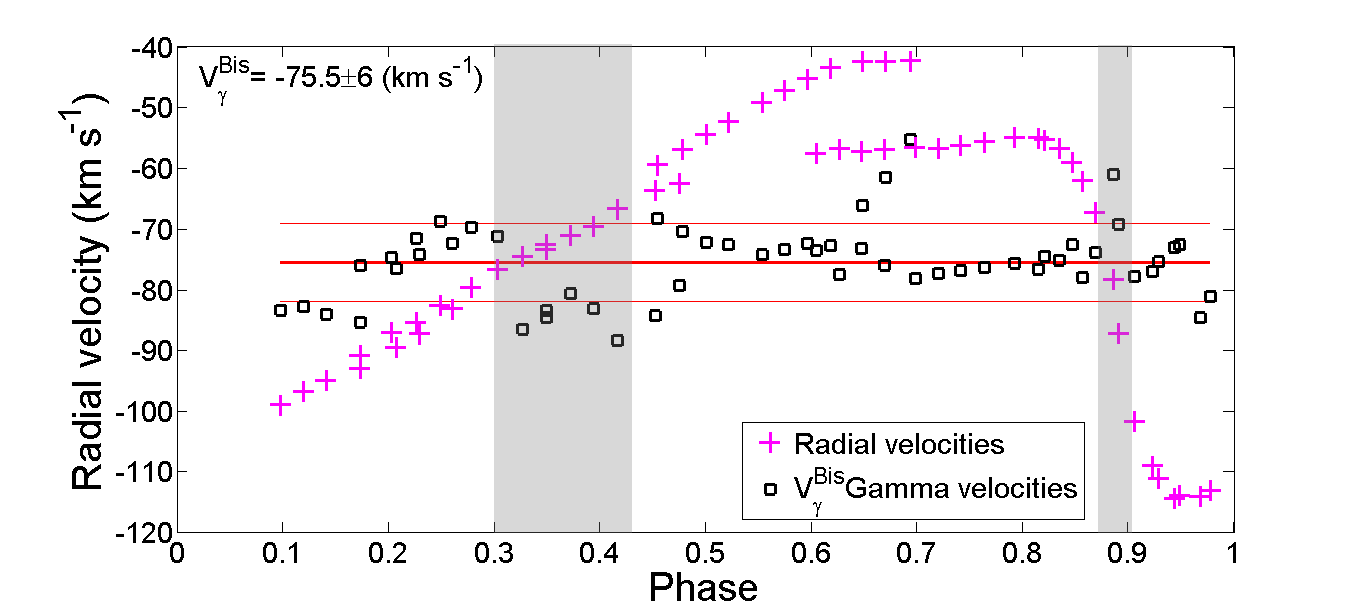}
\caption[]{Top panel: gamma velocities of RR~Lyr for non Blazhko phases derived using the method of bisectors (black squares), inferred from radial velocities measurements (magenta crosses) at different phases. The template radial velocity curve of RR~Lyrae that was using for
Xcor method is drawn as a blue solid line. Bottom panel: the same as previous including the Blazhko phases. Grey regions presumably corresponds to the phases with a low pulsation velocities, that does not affect the line profile
asymmetry.}
\label{fig:rr_lyr}
\end{figure*}

For each spectrum of RR Lyr we computed the LSD-profile as described in
Section~\ref{sec-lsd} and the bisectors as described in Section~\ref{sec:puls},
deriving both $v_{rad}$ and $v_{puls}$ with the same method used for our
program stars. The results are presented in Figure~\ref{fig:rr_lyr}, where the
values of the radial velocity and gamma velocity at different phases are
indicated with magenta crosses and black squares, respectively.
In Table \ref{tab:rrlyr} the individual values of radial and gamma velocities and corresponding values of p-factor are presented.
Moreover, we added to our test case some additional Blazhko phases in order to check the behavior of line profile asymmetries and the performance of the bisectors method in these more difficult phases (see bottom panel of Figure~\ref{fig:rr_lyr}). We can conclude that, even in the most difficult conditions, we obtain $V_{\gamma}^{Bis}$ estimates that are fully consistent with those obtained from other phases.
The individual estimates of $V_{\gamma}^{Bis}$ typically lie within the uncertainty of this average value.
Our analysis shows that Blazhko effect does not effect to an accuracy of the bisectors method. Indeed, the variations in the absolute values of pulsation amplitudes should not affect significantly to the relative difference of pulsational velocity in different layers of the stellar atmosphere that cause spectral line asymmetry. However, possible nonradial modes that occur in Blazhko stars have to be detected by method of bisectors in high resolution and high S/N spectra. The interpretation of nonradial modes in the line profile asymmetry requires a further investigation.

We finally derived an average value of $V_{\gamma}^{Bis} = -73.9 \pm 5.9~km~s^{-1}$. This value is in agreement with literature references obtained with very high quality data, e.g. $V_{\gamma}^{RR Lyr} = -72.0 \pm 0.5~km~s^{-1}$ \citep{chadid_2000}. However, in the phase ranges  0.2--0.3 and 0.85--0.90, deviations up to 18~km~s$^{-1}$ were found (grey shaded areas in Figure~\ref{fig:rr_lyr}).
In these phase ranges, the pulsational velocity is small, and our method is not very sensitive to $|v_{puls}|<5$~km~s$^{-1}$ as it will be discussed in the next subsection. Moreover, strong shocks occur in RR Lyrae's atmosphere at phases 0.85-0.9, contributing to the shape of line profiles, and those phases appear indeed to provide worse performances.

\begin{figure}
\resizebox{\hsize}{!}{\includegraphics[angle=0]{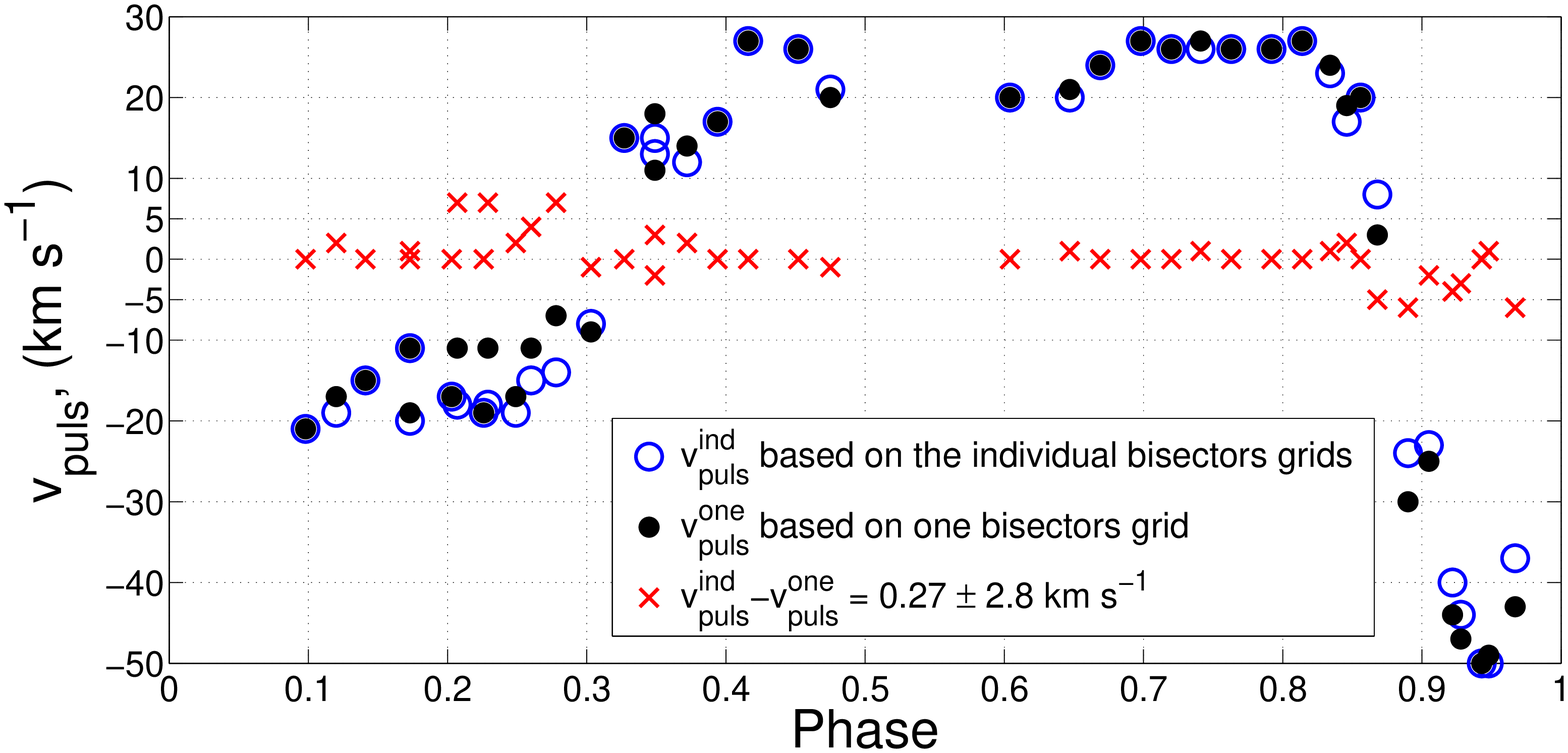}}
\caption[]{The difference between pulsational velocities for each spectrum of RR Lyr derived with the individual sets of bisectors grid and one grid of bisectors number 087 for all observed spectra.}
\label{fig:one_grid}
\end{figure}

In order to test how flexible the bisectors method is, we derived the gamma velocities for RR Lyr with individual sets of bisectors grid for each observation, comparing it with gamma velocities derived using just one bisector grid for all observations. For this test we chose one bisector grid number 087 which corresponds to the phase 0.173 from our RR Lyrae synthetic models (see Table \ref{tab:rrlyr}). In Figure~\ref{fig:one_grid}, we present the pulsational velocities derived for each spectrum of RR Lyr obtained considering individual grids of bisectors and one grid of bisector (ID number 087). The average difference in pulsational velocities along the pulsation curve is relatively small ($< 3~km~s^{-1}$), thus we can conclude that the use of only one bisector grid will not affect much the final gamma velocity estimations.

\begin{figure}
\resizebox{\hsize}{!}{\includegraphics[angle=0]{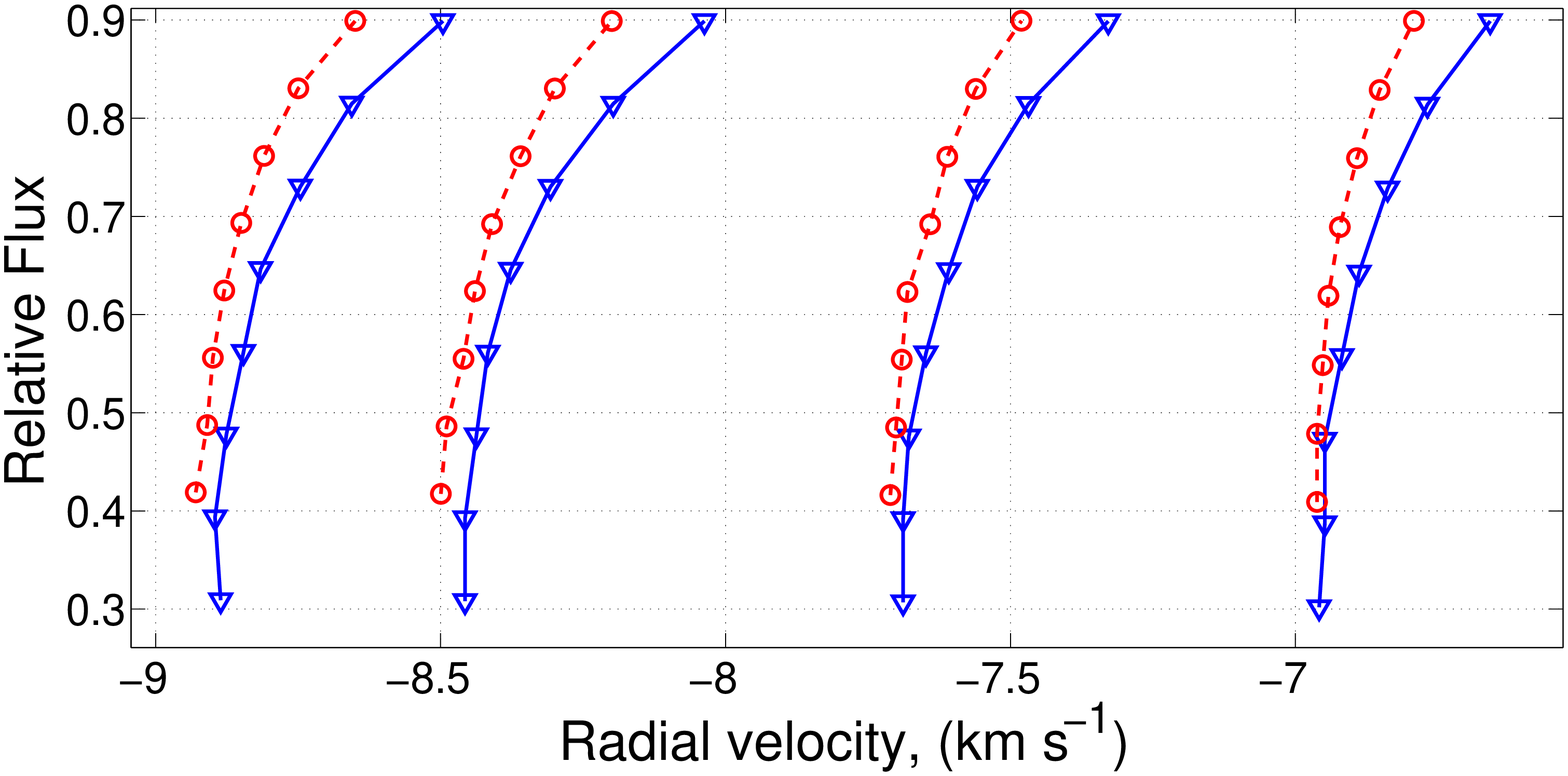}}
\resizebox{\hsize}{!}{\includegraphics[angle=0]{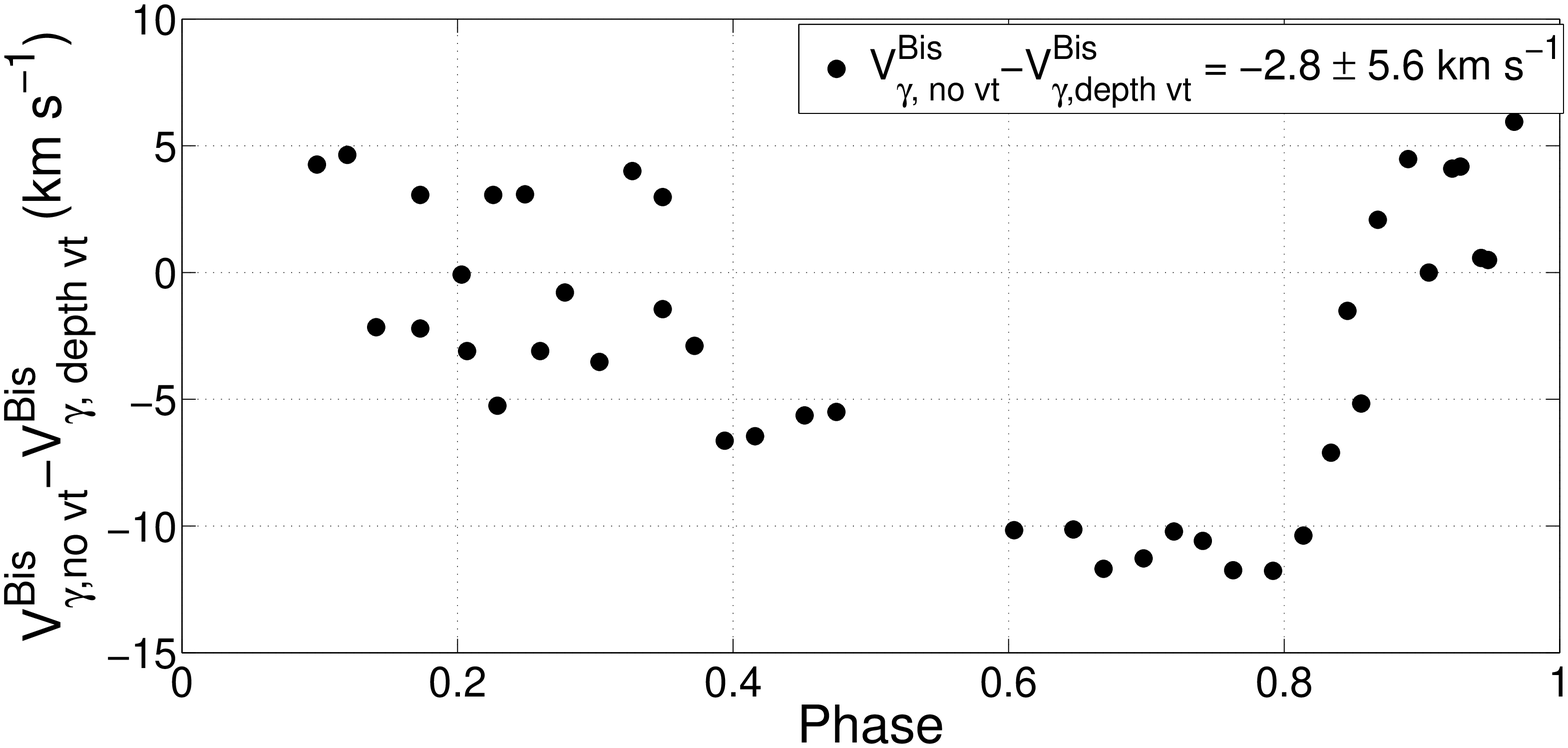}}
\caption[]{Top panel. Comparison of synthetic bisectors computed with the depth-dependent microturbulent velocity model (blue triangles) and with the constant microturbulent velocity $v_{mic}  = 2~km~s^{-1}$  (red circles). Bottom panel. The difference between the gamma velocities of RR Lyr derived with different sets of bisectors grid: with ($V_{\gamma}^{Bis,~depth-dep.~v_{mic}}$) and without depth-dependent $v_{mic}$ stellar atmosphere models ($V_{\gamma, RR Lyr}^{Bis,~constant~v_{mic}}$).}
\label{fig:novt}
\end{figure}

Worth to discuss about one important ingredient, that has a significant impact on the analysis, is the microturbulent velocity; different authors \citep{kolen10,fossati14} found the need of a depth-dependent value of microturbulence, varying with optical depth in the stellar atmosphere. To take the effect into account, we used the trend found by \citet{fossati14} to compute our synthetic spectra. The behaviour of bisector asymmetries computed with depth-dependent $v_{mic}$ stellar atmosphere models and with constant $v_{mic}$ atmosphere models is shown in Figure~\ref{fig:novt}. As a constant $v_{mic}$ value, we adopt $v_{mic}  = 2~km~s^{-1}$, following Paper 1. The difference in residual asymmetries is rather large and, as a result, the difference in derived gamma velocities for each individual observation is significant (bottom panel of Figure~\ref{fig:novt}). It is worth to notice that by using stellar atmosphere models with a depth-dependent microturbulent velocity the residual accuracy of the derived gamma velocity for RR Lyr is much smaller ($V_{\gamma, RR Lyr}^{Bis,~depth-dep.~v_{mic}} = -80.2 \pm 7.8~km~s^{-1}$) than the accuracy of the derived gamma velocity based on models with constant $v_{mic}$ ($V_{\gamma, RR Lyr}^{Bis,~constant~v_{mic}} = -73.9 \pm 5.9~km~s^{-1}$). We can conclude, that the current models of depth-dependent microturbulent velocity are not able to reproduce the real observed behaviour of the line profile asymmetries and the effect of depth dependent microturbulence velocity requires further theoretical investigations. Our analysis suggests that the use of grids of bisectors based on stellar atmospheres with constant $v_{mic}$ velocity should be preferred.

\subsection{The accuracies comparison of the bisectors method and the method of radial velocity curve template.}

\begin{figure}
\resizebox{\hsize}{!}{\includegraphics[angle=0]{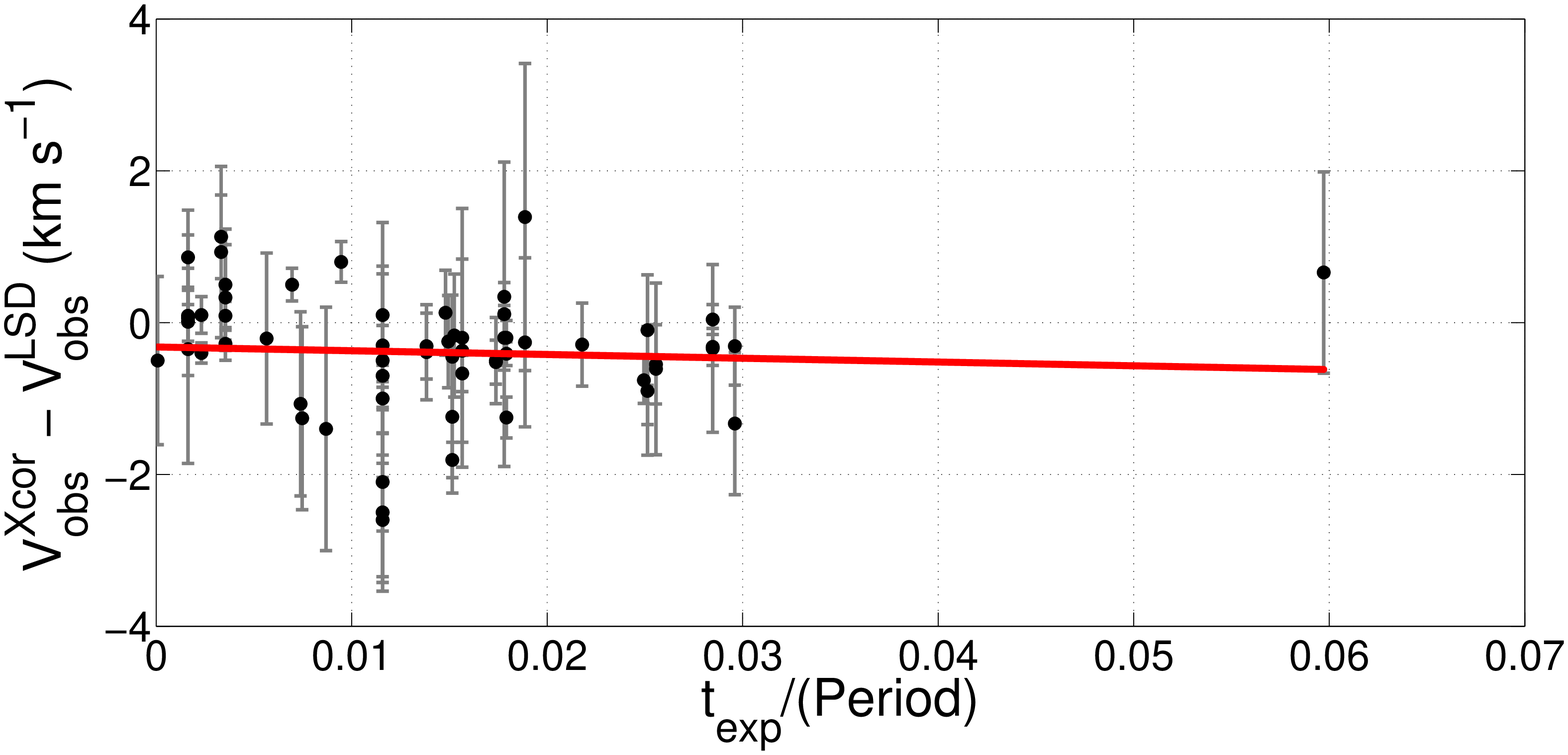}}
\resizebox{\hsize}{!}{\includegraphics[angle=0]{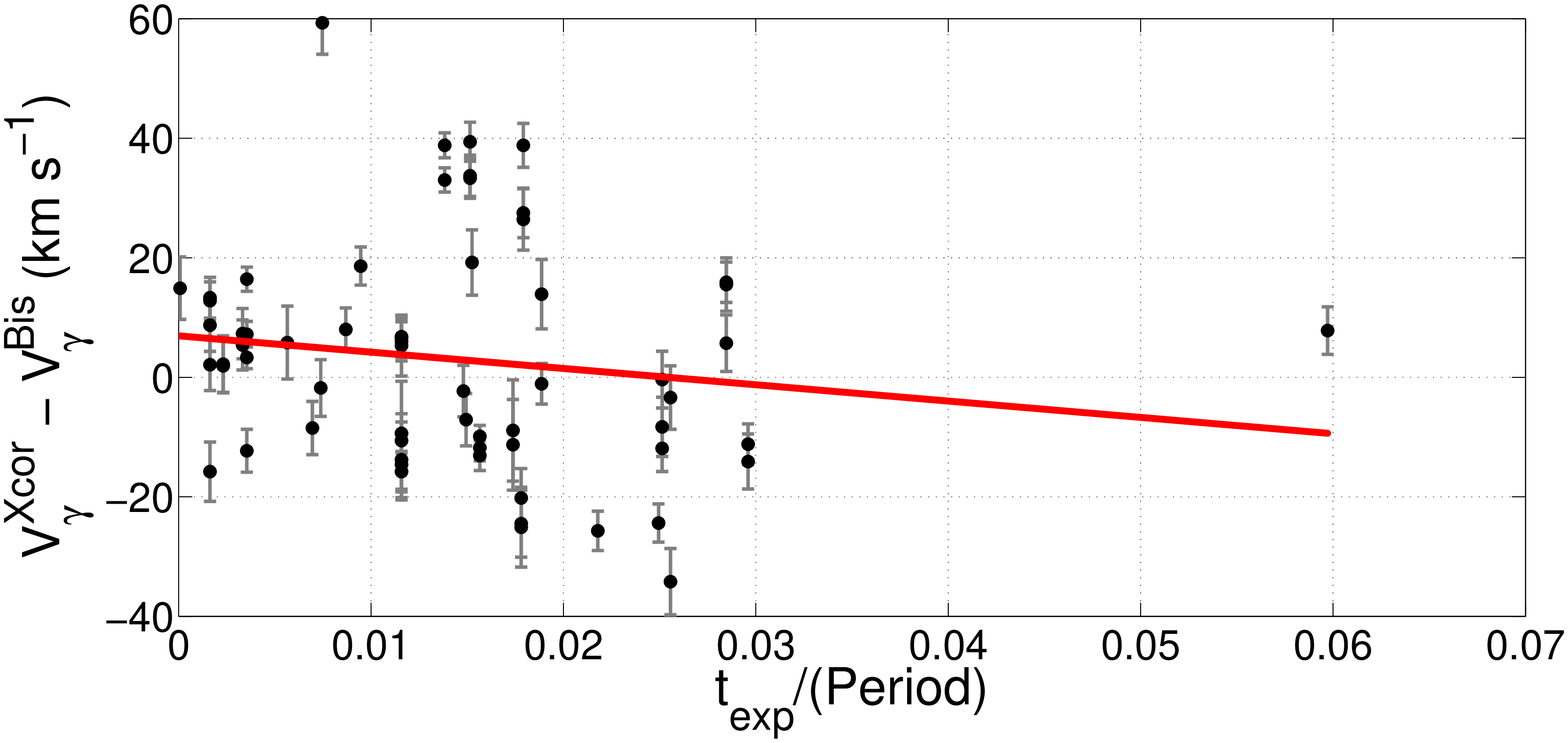}}
\caption[]{Difference between radial velocities (top panel) and gamma
velocities (bottom panel) computed using the cross correlation method and method
of bisectors as a function of exposure time of observed spectra. The red line in each plot corresponds to the linear fits of investigated values.}
\label{fig:obs_texp}
\end{figure}

One source of uncertainty in this method is that observed spectra are usually taken
with finite integration times, that cover a range of phases. The higher the
spectral resolution of the observations, the longer the time needed to obtain a
good S/N ratio, and the larger the {\em phase smearing} effect on the observed
profile. The use of 4 or 8~m class telescopes alleviates the problem, but it is
very difficult to avoid it completely, except for bright stars. The net result is a
general broadening of the observed line profiles, compared to the theoretical ones.
The smearing should cause subtle shifts in the minimum of the LSD profile --- that
should reflect on the derived $v_{obs}$ --- and subtle changes in the line profile
asymmetries --- that should reflect on the derived $v_{puls}$. The size of these
effects depends of course on the phase duration of the exposures. In our sample, we
have exposures as long as 45~min (see Paper~I), so we tested this hypothesis by
comparing the difference between v$_{obs}$ and  V$_{\gamma}$ obtained with the
template curve method with those obtained with the bisectors method, and plotted
them as a function of phase coverage in Figure~\ref{fig:obs_texp}.
As can be seen, there is a hint that the bisectors method might slightly underestimate both velocities for increasing phase coverage, but the smearing effect --- if any --- does not affect much the individual errors of the observed radial velocity measurements. First of all, because when the phase smearing gets larger we do not see any significant trend in increasing the errors of observed and gamma velocities for the program stars. Thus, we can conclude that main sources of the individual errors in observed velocities have rather physical and instrumental (resolution and SNR) nature than the smearing effect. For the better understanding of this effect, it is necessary to perform the same analysis with simulated spectra for the given telescope and spectrograph which requires further studies.


\begin{figure}
\resizebox{\hsize}{!}{\includegraphics[angle=0]{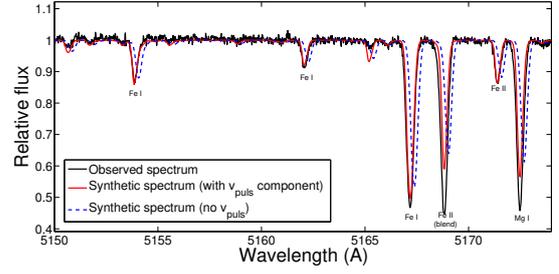}}
\caption[]{A portion of the spectrum of V~Ind observed at phase 0.19 (black solid
line) is superimposed to synthetic spectra with (red solid line) and without
(blue dashed line) a pulsation velocity component. The synthetic spectra assume
stellar parameters derived in Paper~I, i.e. $T_{\mathrm{eff}}$ = 7000 K, $\log g
= 2.3$ dex, $v_{\mathrm{mic}}=1.6$ km~s$^{-1}$. The pulsation velocity determined
for this spectrum is $-17$ km~s$^{-1}$.}
\label{fig:v_ind_sytnh}
\end{figure}

Figure~\ref{fig:v_ind_sytnh} illustrates the quality of our v$_{puls}$
determinations: in spite of the large uncertainties involved in this kind of
measurements, the addition of the v$_{puls}$ component brings the synthetic and
observed spectra in agreement. Moreover, it can be appreciated that the line
shapes are different with and without inclusion of the pulsation velocity.

\begin{figure}
\resizebox{\hsize}{!}{\includegraphics[angle=0]{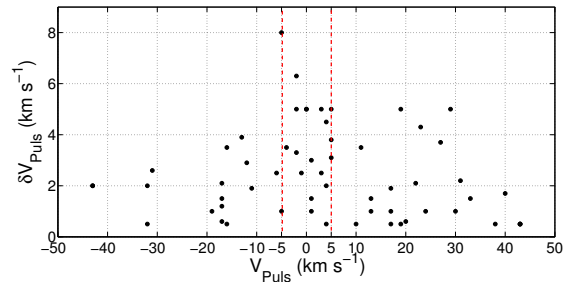}}
\caption[]{The errors on pulsational velocity (see text for details) for each
observed spectrum at different pulsational velocities. }
\label{fig:V_puls_errors}
\end{figure}

The resulting $v_{puls}^{Bis}$ determinations, with their errors $\delta
v_{puls}^{Bis}$, are reported in Table~\ref{tab:ind_lsd}, along with the final
$V_{\gamma}$ and its propagated error, $\delta V_{\gamma}$. The errors on the
pulsational velocity ($\delta v_{puls}^{Bis}$) have been computed as the standard
deviation of measurements $v_{puls}^{Bis}$ obtained from the fit of different
parts of the theoretical and observed bisectors, considering that LSD profiles
can be distorted by blends. Thus, for each observation, we measured the
pulsational velocities by using a different number of LSD profiles layers, namely
we cut the bisectors at the fourth, fifth, sixth, and seventh layer, and derived
a different v$_{puls}$ for each cut. In this way, we tested how reliable
bisectors are in reflecting the asymmetries of LSD profiles. The results are
plotted in Figure~\ref{fig:V_puls_errors}, as a function of pulsational velocity
of each individual exposure. As can be seen, the errors are always below
5~km~s$^{-1}$, with a typical (median) value of 1.5~km~s$^{-1}$,
except for a narrow range of pulsational velocities, from --5 to +5 km~s$^{-1}$, with a median value of 3.5~km~s$^{-1}$. The comparably higher scatter that is obtained for pulsational velocities having smaller values is connected with the intrinsic sensitivity of the method: (i) blends do affect the shape of the LDS profile and (ii) the asymmetry of the profile tends to vanish for small pulsation velocities and consequently the error on its determination increases. This behavior of $\delta v_{puls}^{Bis}$ constrains the range of reliable application of the LSD method and has an impact on the gamma velocity error, $\delta V_{\gamma}$.

\begin{figure}
\includegraphics[width=\columnwidth,angle=0]{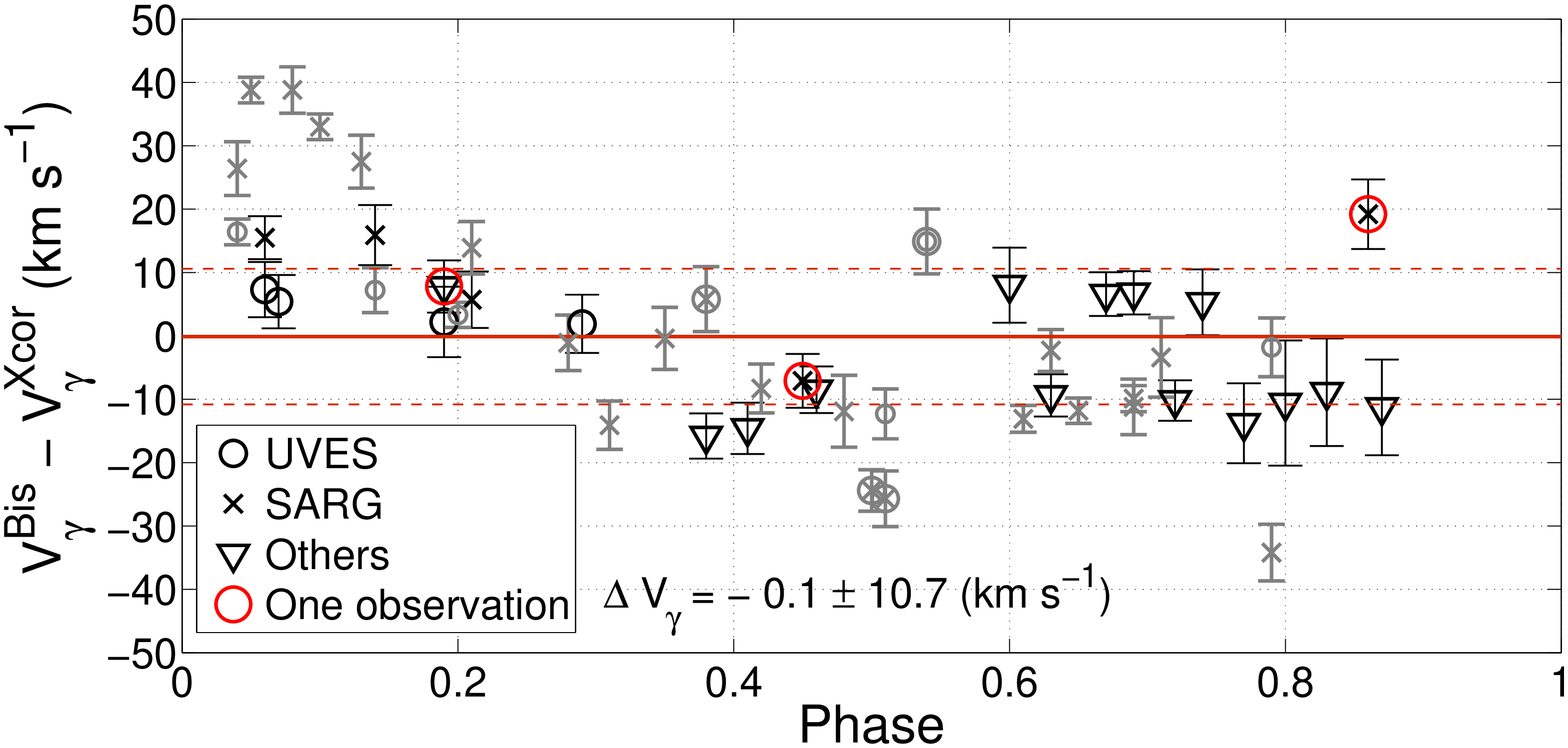}
\caption[]{Difference between gamma velocities derived with the radial
velocity curve template ($V_{\gamma}^{Xcor}$) and with the LSD profile asymmetry
method ($V_{\gamma}^{Bis}$). The observations from the different spectrographs are
labeled respectively, and stars that have only one epoch spectrum are also marked.
Stars marked in black are those having reliable observed radial velocity curves in
the literature (e.g., from the Baade-Wesselink method), while stars marked in grey have $V_{\gamma}^{Xcor}$ derived from less reliable templates.}
\label{fig:gamma_xcor-bis}
\end{figure}

Figure~\ref{fig:gamma_xcor-bis} compares the final V$_{\gamma}$ obtained with the
classical and bisector methods. As can be seen, the stars that do not possess reliable
observed curves in the literature (for example from the Baade-Wesselink method)
have a larger difference, that is mostly caused by uncertainties in the classical template fitting method.
If we limit the comparison to the stars having reliable template radial velocity curves, we can see an overall agreement at the level of $\Delta$V$_{\gamma}$=--0.1$\pm$10.7~km~s$^{-1}$.

We also note that the most discrepant stars (marked in grey in Figure~\ref{fig:gamma_xcor-bis}) are observed with SARG, and generally have lower S/N ratios and a larger phase coverage (smearing), because it is mounted on a 4~m-class telescope, while the UVES spectra produce less scattered results.

\begin{figure}
\includegraphics[width=\columnwidth,angle=0]{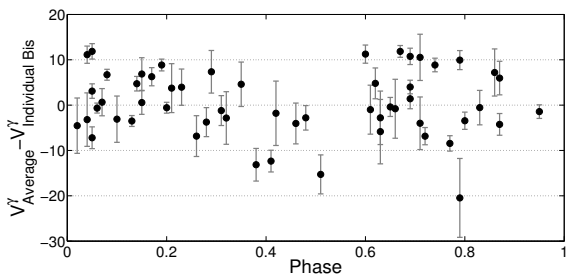}
\caption[]{The difference of $V_{\gamma}^{Bis}$ of individual exposures with the average $V_{\gamma}$ for each star as a function of phase.}
\label{fig:gamma-bis}
\end{figure}

Another way to study {\em a posteriori} the error on the whole method consists
in evaluating the spread on repeated $V_{\gamma}$ measurements for the same
star, when more than one determination was available. We found a typical spread of --10 to +10 km~s$^{-1}$, with the maximum variations corresponding generally to the phase ranges 0.4--0.5.
Figure~\ref{fig:gamma-bis} shows the difference of $V_{\gamma}$ measurements obtained from individual exposures with the average $V_{\gamma}$ of each star, as a function of phase. No systematic trends are apparent and the spread is generally consistent with the estimated errors.

\subsection{Projection factor}
\label{sec-dark}

When deriving radial velocities from absorption line profiles, either with
cross-correlation or with LSD profiles, the information is implicitly integrated
over the stellar surface and along the radius of the pulsating star. Such
integration is affected by limb darkening, across the stellar surface, and by
radial velocity gradients with depth in the atmosphere of the star, where lines
form. Both effects depend on the pulsation phase \citep{marengo03}, and therefore
they are generally modeled as:

\[
p = v_{puls}/(v_{rad}-V_{\gamma})  \qquad (2)
\]

\noindent also known as projection factor, or the factor by which one should
multiply the observed radial velocity (in the line of sight) to
obtain the intrinsic pulsational velocity of the star (along the radial
direction), corrected by projection effects.

\begin{figure}
\resizebox{\hsize}{!}{\includegraphics[angle=0]{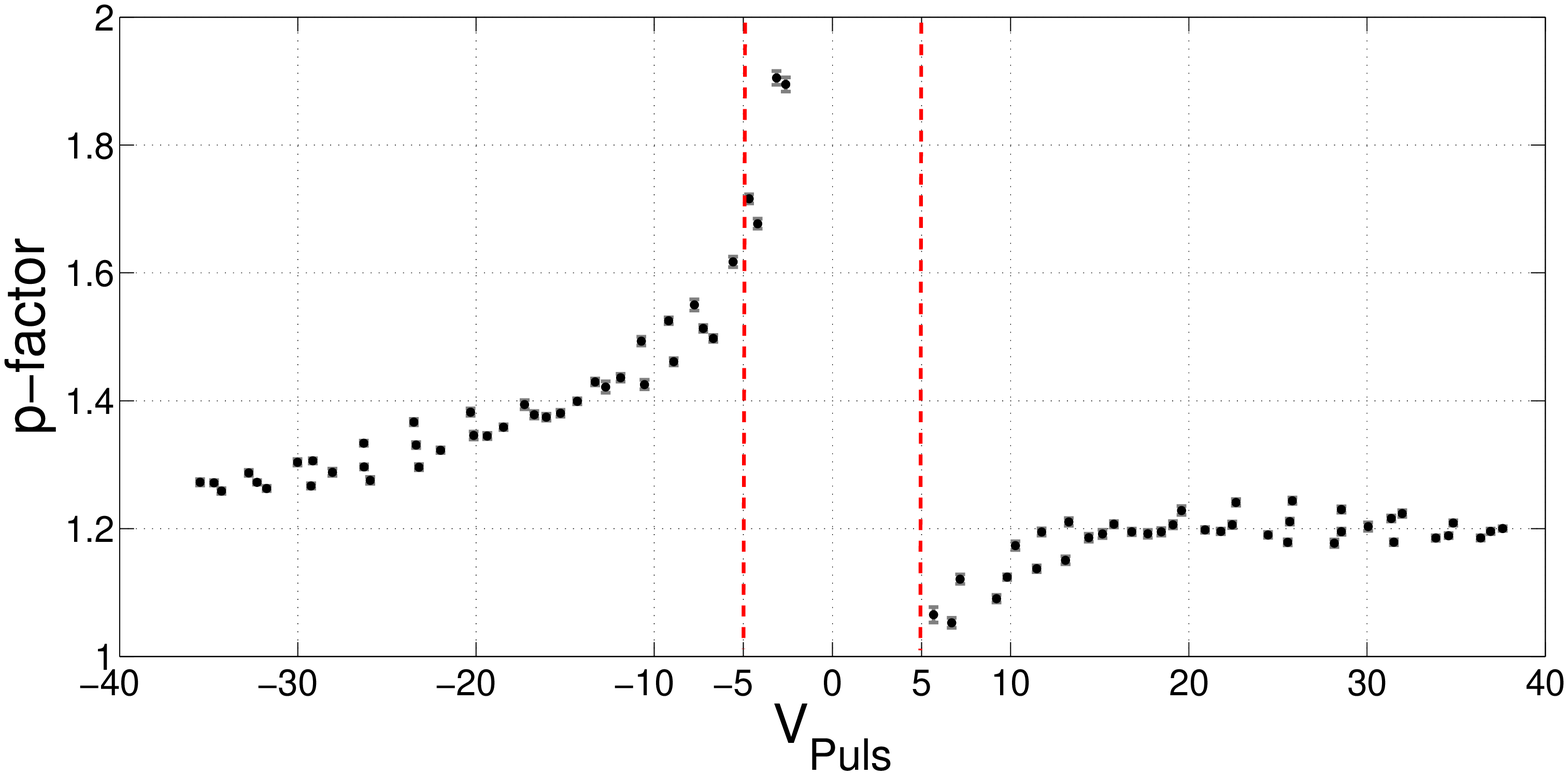}}
\resizebox{\hsize}{!}{\includegraphics[angle=0]{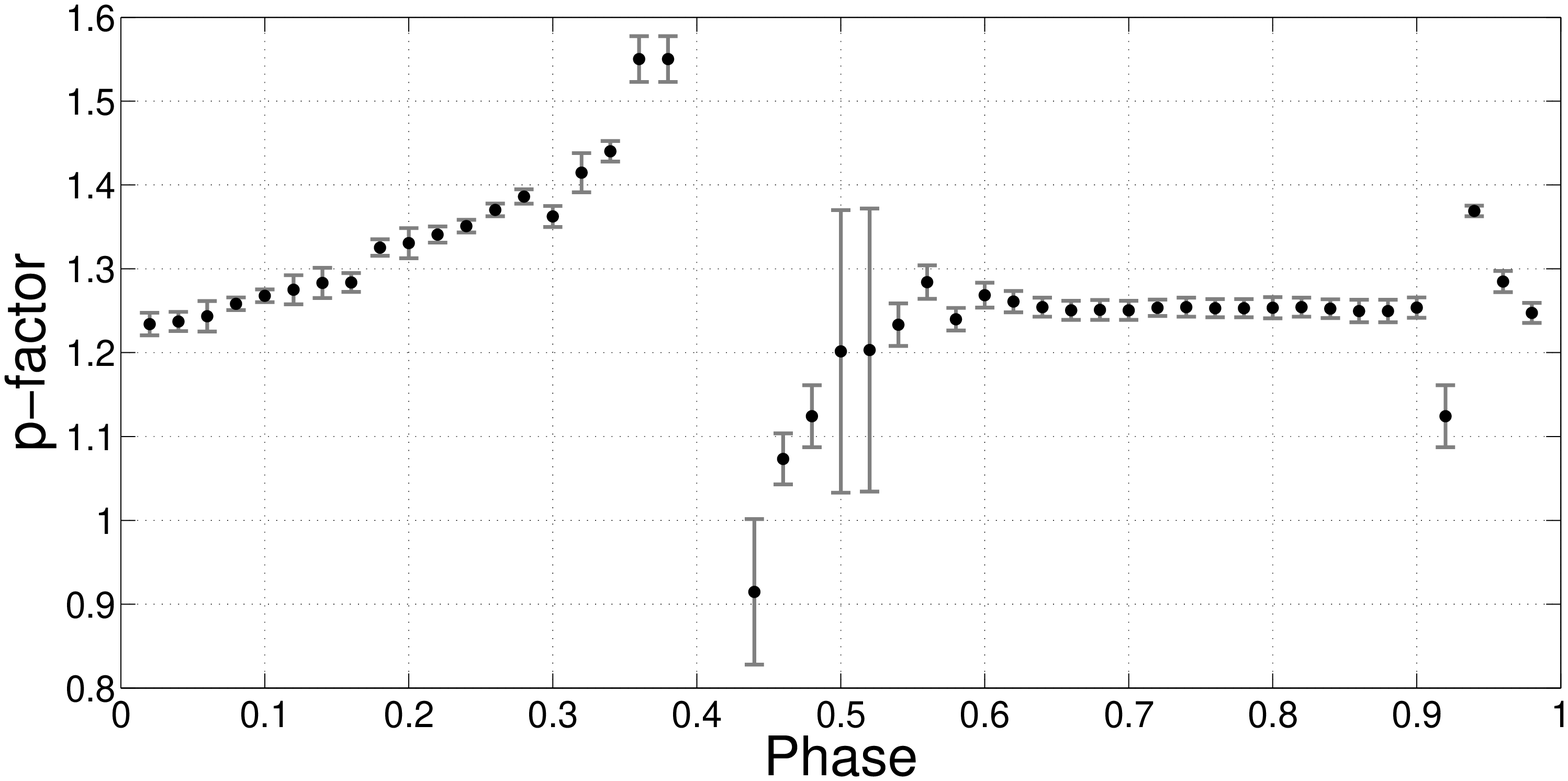}}
\caption[]{Model behavior of p-factor with different pulsation velocity (top
panel) and different phases (bottom panel). We plotted the average values of the projection factor for each value of the modelled pulsational velocity, over all 41 bisector grids (see Appendix~\ref{sec-lib}). The
vertical red dashed lines in the top panel mark the limits of
applicability of our method (from --5 to +5 km~s$^{-1}$, see also
Figure~\ref{fig:V_puls_errors}), that roughly correspond to the phase range
0.35-0.50.}
\label{fig:pfactor}
\end{figure}

\begin{figure}
\resizebox{\hsize}{!}{\includegraphics[angle=0]{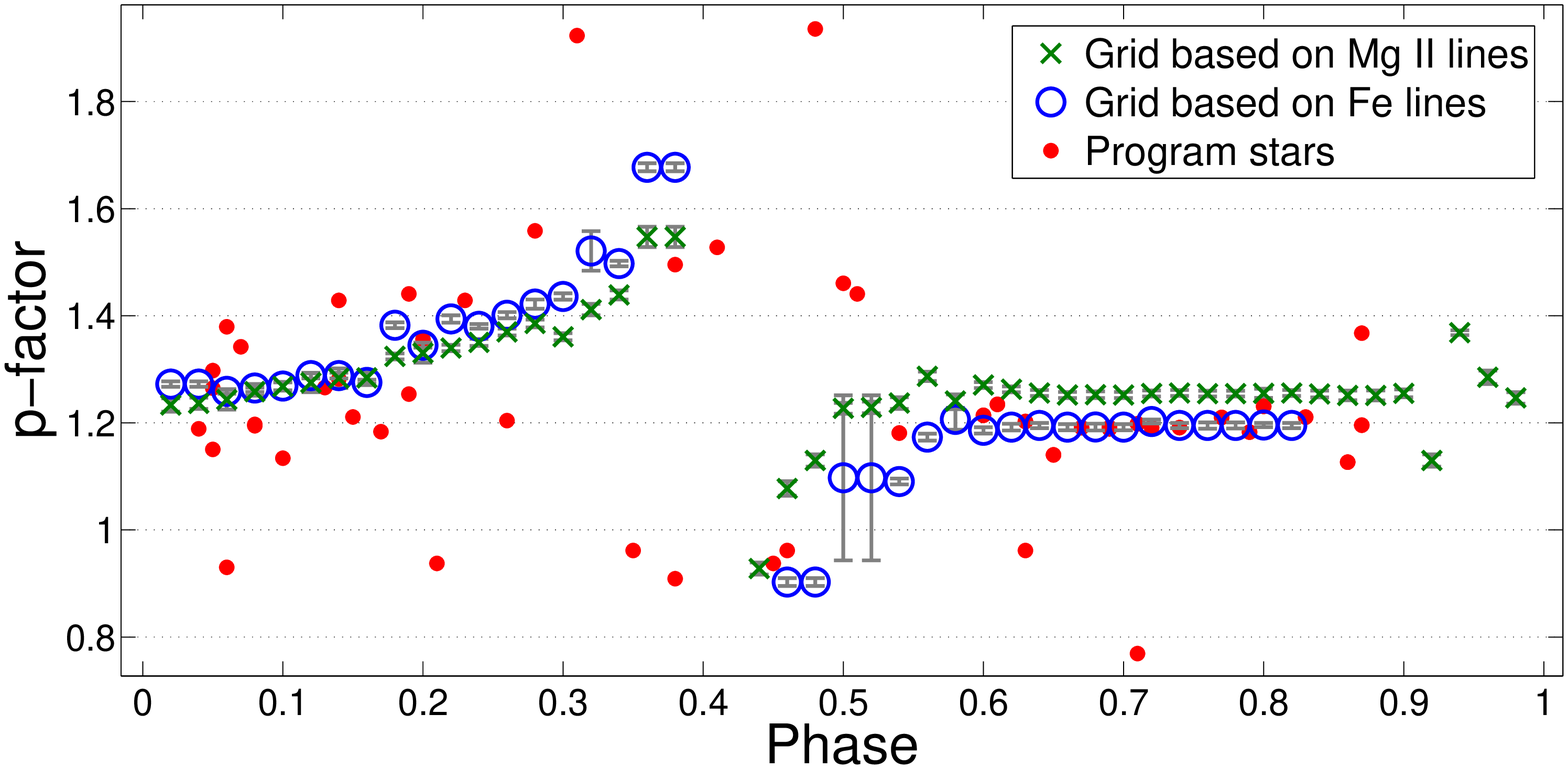}}
\caption[]{Observed behavior of p-factor with different phases for a sample of
program spectra (red dots). As a reference the theoretical values of p-factor are labeled by blue circles and green crosses, which are based on the LSD profiles of synthetic spectra for all lines in the range 5100--5400~\AA (blue circles) and the Mg I triplet lines only (green crosses).}
\label{fig:observed_pfactor} \end{figure}

The projection factor becomes very important in the Baade-Wesselink method,
where a relatively small error in pulsation velocity may lead to a large error
in distance. The first works by \citet{eddington} or \citet{carroll} lead to an
estimate of $p=1.41$ that was used in many Baade-Wesselink studies of Cepheids,
to correct for limb darkening and radial velocity expansion effects on the
derived distances. Later, different values were proposed, ranging from 1.2 to
1.4 \citep[see][for a review]{Nardetto2014_rev}. For the closest stars, it is
possible to measure the projection factor  directly, with the help of
interferometric observations \citep[e.g.][]{Breitfelder}. An investigation with
realistic hydrodynamical models of $\delta$~Cep by \citet[][see also the subsequent
work by \citealt{Nardetto2013}, \citealt{Nardetto2014} and \citealt{Nardetto2017}] {Nardetto2004} showed
that projection effects can be corrected with residuals below
$\simeq$1~km~s$^{-1}$ and that the appropriate projection factor depends on
the method used to derive the projected pulsational velocity.

In our case, the projection factor was computed using the bisectors method \citep[an application to Cepheids can be
found in][]{sabey95}. We modeled the theoretical line profiles by taking into
account the geometrical effect (limb darkening) for each model that we have used for calculations in
our library of bisectors. The particularity of our approach is that we are using the average profiles of different spectral lines (via LSD technique) in order to investigate the clear geometrical effect of radial pulsations.

More in detail, we used the 41 existed models for RR Lyrae over all pulsation phases from \citet{fossati14}. See Appendix~\ref{sec-lib} for more details. For each model, corresponding to a different phase, we computed a set of synthetic spectra with pulsation velocities ranging from -50~km~s$^{-1}$ to +50~km~s$^{-1}$ with a step of 1~km~s$^{-1}$, in the same wavelength range used for our $v_{obs}$ measurements, 5100--5400~$\AA$ (see Figure~\ref{fig:template_lsd}). We then computed $v_{obs}$ and v$_{puls}$, as done on the observed spectra in our sample. We could then compute the projection factor using equations (1) and (2). In Figure \ref{fig:pfactor}, we illustrate the behavior of our modeled p-factor --- obtained from our grids of model bisectors (Appendix~\ref{sec-lib}) --- versus pulsational velocity and phase.
The top panel clearly shows the limits of applicability of the bisectors method, which looses sensitivity in the -5~km~s$^{-1}$ to +5~km~s$^{-1}$ range, as discussed also previously.
As a comparison, we repeated the same analysis using just the Mg I triplet lines (5167.321, 5172.684, 5183.604 $\AA$) instead of the all prominent lines from our initial linelist. In this case we built the synthetic grid of bisectors, and we derived the observed ones of our stellar sample, based only on these three lines. The Mg triplet is a prominent feature in the spectra of RR Lyraes, and it allowed us to test our p-factor modelization under different conditions (less numerous but stronger lines, see Figure~\ref{fig:observed_pfactor}).

In order to investigate the behavior of the projection factor on the sample of our real stars, we repeated the same procedure to derive the p-factor on the observed LSD profiles for the whole sample of stars. We can also plot the {\em observed} p-factor for each of the analyzed spectra, illustrated in
Figure~\ref{fig:observed_pfactor}. Considering that we used the same bisectors library, we would expect to obtain roughly the theoretical behavior illustrated in the lower panel of Figure~\ref{fig:pfactor}, modulated by the fitting
uncertainties. While globally this is true, there are several outliers, that we can see in Figure~\ref{fig:observed_pfactor}, displaying in some cases totally unrealistic values of the p-factor, for example $p$$<$1 and $p$$>$1.6. This is most likely caused by measurement uncertainties in the derived v$_{puls}$, caused by line smearing, low S/N, blends, and other observational effects, and discussed in the previous section. But the overall behavior of p-factors derived from the observed profiles of our sample stars appears to agree well with that obtained from the models based on RR Lyrae.
This confirms the overall self-consistency of our analysis. The comparisons of p-factors derived from our whole line list and from the Mg triplet shows a small offset that varies with phase (up to 0.07). We conclude that line selection can have a noticeable impact on the actual p-value derivation. When the offset is compared with the v$_{obs}$ spread, however, it becomes largely irrelevant.

It should be noted, in the framework of our analysis we assume that all lines even with a different individual depths are behaving in the same way -- i.e. the shape of all spectral features is the same at given pulsational velocity. This is not a fully correct assumption, however, the primary goal of this paper is to develop a method which can give reliable estimations of the gamma- and pulsational velocities just from one spectral observation. For this reason, using a few prominent lines or the list of many weak lines is equally good for deriving investigated velocities.

\begin{table}
\caption{Final results for the program stars, obtained with both the
cross-correlation and the bisectors methods.}
\centering
\label{tab:results}
\begin{tabular}{lrrrr}
\hline
Star     & $V_{\gamma}^{Xcor}$ & $\delta V_{\gamma}^{Xcor}$ & $V_{\gamma}^{Bis}$ & $\delta V_{\gamma}^{Bis}$ \\
         & (km~s$^{-1}$)       & (km~s$^{-1}$)              & (km~s$^{-1}$)      & (km~s$^{-1}$) \\
\hline
DR And   & --119  &  9 & --110  &  4 \\
X Ari*   &  --36  &  5 &  --44  &  6 \\
TW Boo   & --100  &  2 &  --89  &  2 \\
TW Cap*  &  --72  &  6 &  --87  &  6 \\
RX Cet*  &  --94  &  5 &  --68  &  6 \\
U Com    &   --7  &  4 &  --19  &  6 \\
RV CrB   & --140  &  4 & --133  &  5 \\
UZ CVn   &  --18  &  4 &  --49  &  6 \\
SW CVn   &    21  &  7 &    14  &  6 \\
AE Dra   & --276  &  2 & --311  &  5 \\
BK Eri   &    97  &  4 &    93  & 11 \\
UY Eri*  &   152  &  6 &   146  &  7 \\
SZ Gem*  &   322  &  5 &   346  &  6 \\
VX Her   & --371  &  5 & --352  & 10 \\
DH Hya   &   334  &  5 &   348  & 18 \\
V Ind    &   197  &  3 &   200  &  9 \\
SS Leo   &   164  &  5 &   157  &  2 \\
V716 Oph*& --311  &  6 & --370  &  6 \\
VW Scl   &    40  &  4 &    36  & 13 \\
BK Tuc   &   170  &  5 &   134  &  7 \\
TU UMa*  &    99  &  5 &   106  &  7 \\
RV UMa   & --198  &  5 & --175  &  6 \\
UV Vir*  &    99  &  5 &    81  &  6 \\
\hline
\end{tabular}
\medskip \begin{flushleft} \emph{Note.} * For these stars only one spectrum
was obtained. \end{flushleft}
\end{table}

\section{Discussion}

The final velocity measurements for each star, measured with the two
methods, were obtained as averages of the single-epoch
gamma determinations listed in Tables~\ref{tab:ind_ccf} and
\ref{tab:ind_lsd}, respectively, and they are listed in Table~\ref{tab:results}
along with their errors. The comparison between the two methods was already
discussed in Section~\ref{sec-new} (see Figures~\ref{fig:obs_texp} and
\ref{fig:gamma_xcor-bis}). Here, we will compare our results with the
available literature. We will also discuss more in depth the
uncertainties and applicability of the bisectors method to the derivation of
gamma velocities for pulsating stars.

\begin{table}
\caption{Summary of literature results on the program stars, for both the
observed heliocentric radial velocity estimates ($v_{rad}$) and the gamma
velocities (V$_{\gamma}$), along with their errors, when available.}
\centering
\label{tab:literature}
\begin{tabular}{l@{}r@{\hskip 5pt}rr@{\hskip 5pt}r@{\hskip 5pt}rr}
\hline
Star     & $v_{rad}^{Lit}$ & $\delta v_{rad}^{Lit}$ & Ref & $V_{\gamma}^{Lit}$ & $\delta V_{\gamma}^{Lit}$ & Ref\\
         & (km s$^{-1}$)  &  (km~s$^{-1}$)  && (km~s$^{-1}$) & (km~s$^{-1}$) &\\
\hline
DR And   &  --81 & 30 & (1) & --103.6 & 2.0  & (3) \\
X Ari*   &  --36 &  1 & (1) &  --41.6 &  3   & (4) \\
TW Boo   &  --99 &  1 & (1) &  --99   &  5   & (5) \\
TW Cap   &  --32 &  1 & (1) &  ...    & ...  & \\
RX Cet   &  --57 &  2 & (1) &  ...    & ...  & \\
U Com*   &  --22 &  3 & (1) &  --18   &  3   & (6) \\
RV CrB   & --125 &  5 & (1) & --125   &  5   & (7) \\
UZ CVn*  &  --28 &  3 & (1) &  --27   & 10   & (6) \\
SW CVn   &  --18 & 21 & (1) &  --18   & 21   & (8) \\
AE Dra   & --243 & 30 & (1) & --243   & 30   & (8) \\
BK Eri   &   141 & 10 & (1) &   141   & 10   & (7) \\
UY Eri   &   171 &  1 & (1) & ...     & ...  & \\
SZ Gem   &   326 &  4 & (1) &   305   & 15   & (5) \\
VX Her*  & --377 &  3 & (1) & --361.5 &  5   & (4) \\
DH Hya   &   355 &  8 & (1) &   355   &  8   & (7) \\
V Ind*   &   202 &  2 & (1) &   202.5 &  2   & (9) \\
SS Leo*  &   144 &  7 & (2) &   162.5 &  6.8 & (10)\\
V716 Oph & --230 & 30 & (1) & --230   &  5   & (11)\\
VW Scl   &    53 & 10 & (1) &    53   & 10   & (12)\\
BK Tuc   &   121 & 14 & (1) &   121   & 14   & (7) \\
TU UMa*  &    89 &  2 & (1) &   101   &  3   & (13)\\
RV UMa   & --185 &  1 & (1) & --185   &  1   & (7) \\
UV Vir   &    99 & 11 & (1) &    99   & 11   & (7) \\
\hline
\end{tabular}
\medskip \begin{flushleft} \emph{Notes.} * These stars were studied with the
Baade-Wesselink method. Literature references: (1) \citet{beers00}; (2)
\citet{rave}; (3) \citet{jef07}; (4) \citet{nemec13}; (5) \citet{hawley85}; (6)
\citet{fern97}; (7) \citet{dambis09}; (8) \citet{layden94}; (9) \citet{vind};
(10) \citet{carrillo95}; (11) \citet{mac94}; (12) \citet{solano97}; (13)
\citet{preston67}.
\end{flushleft}
\end{table}

\subsection{Literature comparisons}

\begin{figure*}
\includegraphics[scale=0.7,angle=0]{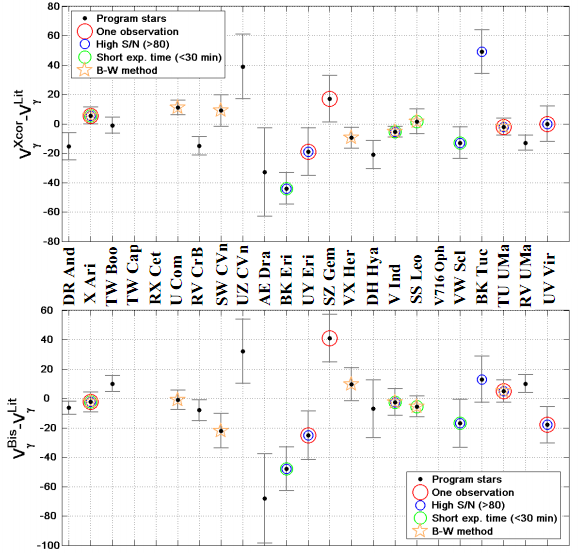}
\caption[]{Differences between our measurements and literature ones. Top panel:
difference between gamma velocities from cross-correlation method and
template curve fitting (V$_{\gamma}^{Xcor}$) with the literature
(V$_{\gamma}^{Lit}$). Bottom panel: difference between the gamma velocity
from LSD profiles and bisectors (V$_{\gamma}^{LSD}$) with the literature.
We marked the stars by different colors taking into account some relevant characteristics of their spectra: the stars that have only one measurement (red), short exposure times (green), high SNR (blue), and Baade-Wesselink literature estimates (orange).}
\label{fig:xcor-bis}
\end{figure*}

We report in Table~\ref{tab:literature} our literature search for radial and
gamma velocity measurements of the program stars (V$_{\gamma}^{Lit}$). We compare the gamma
velocities obtained in the literature with our measurements obtained with both
methods in Figure~\ref{fig:xcor-bis}. As can be seen, both the classical method (top
panel) and the  bisector method (bottom panel) produce results that are
generally compatible with the literature, within the quoted uncertainties, with a
few exceptions.

We note that for the most discrepant cases, like AE Dra, UZ CVn, or BK Eri, the
discrepancy with the literature persists regardless of the method employed,
suggesting that maybe there is a problem with those specific stars.
The spectra of SW~CVn and RV CrB have low S/N ratio ($\le$50) and they were excluded from
the abundance analysis in Paper~I, however, we included them in this work, because the S/N ratio was sufficient for radial velocity analysis. These stars were included here because the S/N ratio was sufficient to produce reliable radial velocity estimates, and indeed they agree with the literature with both methods. The spectra of AE~Dra have good S/N ratios but they suffer from some phase smearing because they were observed for 30 and 45 minutes and this is reflected in their large error bars, which make them only marginally incompatible with the literature. The spectra of BK~Eri have quite high S/N and short exposure times, so we suspect that in that case the problem might lie in the literature measurements. Finally, there are two stars that show marginally discrepant values with the literature, and from one method to the other: SZ~Gem and BK~Tuc. SZ~Gem has only
one spectrum and this can pose a problem of non reliable template curve fitting or non sensitive regime of pulsation phases in the bisectors method -- between --5~km~s$^{-1}$ and +5~km~s$^{-1}$. Indeed, the derived pulsational velocity for this spectrum is relatively small 13~km~s$^{-1}$ at phase 0.50. BK~Tuc was observed in three exposures of 30~min each, which might introduce some phase smearing of the line profiles.

If we exclude the three most problematic stars, both methods compare quite well with
the literature, with the following weighted average differences:
$\Delta$V$_{\gamma}^{Xcor}$=--3$\pm$20~km~s$^{-1}$ and
$\Delta$V$_{\gamma}^{Bis}$=--3$\pm$30~km~s$^{-1}$, where the differences are
computed as our measurements minus the literature ones. The observed spreads are
marginally incompatible with the typically quoted uncertainties, which are of 5, 12,
and 7~km~s$^{-1}$ for the cross-correlation method, the bisectors method, and the
literature collection, respectively. The uncertainties quoted for the different methods are rather compatible with each
other, suggesting that there must be some error source that has been unaccounted for
in {\em all} of the methods. If so, the unaccounted-for error source amounts to
10--20~km~s$^{-1}$ in each method. Systematic errors of this amplitude are easy to accommodate when one applies the template curve fitting to just a few measurements as in our case. These errors are also easy to accommodate, at least for the noisiest
spectra, into the bisectors method sensitivity. On the other hand, we are comparing
with literature results that are derived with several different methods and data
samples, so the inhomogeneity of the comparison measurements certainly plays an
important role. Explanation of this general spread in comparison of gamma velocities with the literature values, likely can be found in the fact that the stars for which the B-W analysis was performed have a difference in this comparison less than or equal to 10 ~km~s$^{-1}$ in both V$_{\gamma}^{Xcor}$ - V$_{\gamma}^{Lit}$ and V$_{\gamma}^{LSD}$ - V$_{\gamma}^{Lit}$ values. It is a rather small difference, and the reason for this is actually that these stars have reliable radial velocity curves based on the B-W method. As a result the V$_{\gamma}^{Xcor}$ are more reliable as well, because the light curve analysis was based on the same radial velocity curves as in the B-W method. Moreover, in Figure \ref{fig:gamma_xcor-bis} we can see the same behavior of difference V$_{\gamma}^{Xcor}$ - V$_{\gamma}^{LSD}$  -- the stars for which the B-W method was applied in the literature, have less scatter in the gamma velocity.

We can conclude, after considering the literature comparison with stars having reliable radial velocity curves, that the bisector's method performs at least as well as the more traditional template curve technique.

\section{Conclusions}

We perform a systematic study of the radial velocities for a sample
of 23 variable stars -- mostly RRab, with two RRc and three W~Vir -- using
high-resolution spectra from both proprietary (UVES and SARG) and archival (UVES,
HARPS, FEROS, APO) data sets.

For each star, we derive the radial velocity using two independent methods: the
cross-correlation approach and the LSD profile technique. The errors of
individual estimates of radial velocity in both methods are on the level
$\pm~2~km~s^{-1}$, or less. The determination of gamma velocities was also performed with two methods, the first based on the classical radial velocity curve template technique and the second on the line profile asymmetries, referred to as the bisectors method (see Section \ref{sec:puls}). With the help of LSD technique we provide and test a method
for determining the pulsation and gamma velocity of pulsating variable
stars when only scarce observations taken at random phases are available. We also computed a grid of synthetic bisectors for each of 41 phases (see Appendix A), that is made available electronically.
Our public library can be used in general case to make an
assumptions about velocity of barycenter of pulsating stars, in best case within
error $\pm~5~km~s^{-1}$.

In summary this work, the presented method of deriving gamma velocities of RR Lyraes has the next milestones:

\begin{enumerate}
\item the method is working well for the radially pulsating stars: RRab Lyraes, classical Cepheids; to quantify the level of line profile asymmetry, we used the average profile of different spectral lines (the LSD profile);
\item the method is phase sensitive, the best accuracy of derived at phases 0 -- 0.2 and 0.5 -- 0.85; at these phases the pulsation velocity is large, thus the line profile asymmetry is more easy to identify;
\item the shock phases during the pulsation cycle in RRab stars do affect the accuracy of the method mostly on phases 0.85 -- 0.95, when the line profile geometry changes dramatically;
\item the method is working with both low- and high- S/N spectra, the spectra resolution at which we applied the method was about $R\approx30000$;
\item the major advantage is that it can derive the gamma velocity of a star just from one observation, or from observations taken at unknown phase, however, the accuracy of the method depends on many factors, we estimated the average error of determined pulsational velocity is $\delta$$v_{puls}$ $\approx$ 3.5~km~s$^{-1}$ and the residual error of gamma velocity is $\delta$V$_{\gamma}^{Bis}$ $\approx$ 10~km~s$^{-1}$.

\end{enumerate}

\section*{Acknowledgments}
We thank anonymous referee for the helpful comments which significantly improved the paper. N.~Britavskiy acknowledges partial support under MINECO projects AYA2015-68012-C2-1-P and SEV 2015-0548. Also N. Britavskiy warmly thanks Gisella Clementini and Carla Cacciari for useful recommendations during work on this paper, and all the INAF\,--\,Bologna Observatory, where most of this work was carried out, for the hospitality during the grant stay. D.~Romano
acknowledges financial support from PRIN MIUR 2010--2011, project ``The chemical
and dynamical evolution of the Milky Way and Local Group galaxies", prot.
2010LYSN2T. V.~Tsymbal acknowledges the partially support by RFBR, research
project No.15-52-12371. In this work we made extensive use of the NASA ADS
abstract service, of the Strasbourg CDS database, and of the atomic data compiled
in the VALD data base.

\appendix

\section[]{Grids of synthetic bisectors.}
\label{sec-lib}

We publish a grid of synthetic bisectors, that can be used to compute the
pulsational velocity of any observed spectrum of radial pulsating variables of the
RRab~Lyrae type and Cepheids with the method described in Section~\ref{sec-new}.
Due to the small number of RRc variables (only 2) in our sample we could not make statistically significant conclusions about the method application to this type of variables.
The grid contains one set of bisectors (from $-50$ to $+50~km~s^{-1}$ with a step $1~km~s^{-1}$ of pulsational velocity variation) for each of the 41 different phases computed on the basis of the individual atmospheric models along the pulsation cycle of RR Lyr. Each individual atmospheric model was based on the stellar parameters derived in \cite{fossati14} and phase coverage is presented in the Table \ref{tab:rrlyr}. 
The IDs of each bisector grid correspond to the IDs of the atmospheric models from \cite{fossati14}. 

In order to build the grid of bisectors, we were following the next steps. For each of the 41 selected phases we computed a set of synthetic spectra in the same wavelength range used for the radial velocity
measurements (5100--5400~\AA) with the STARSP-SynthV code \citep{T96}, based on the ATLAS9 model atmospheres. We used the last modification of the code that for the construction of line profiles took into account the stellar radial pulsations, individual chemical abundances, and depth-dependent microturbulent velocity for a given star. For the calculations, we adopt a constant microturbulent velocity of $2~km~s^{-1}$ . However, as discussed at the end of subsection 4.3, the choice of one bisector grid for the analysis of all observed spectra will have no effect on the accuracy of the final gamma velocity measurements. For each synthetic spectrum, we computed the LSD profile and its bisector, using exactly the same method adopted for the observed spectra (Section~\ref{sec-new}). The resulting sets of bisectors serve us as a final library of bisectors.



\begin{table*}
\caption{The grids of synthetic bisectors. The names of the files correspond to the spectrum ID of that phase by \citet{fossati14}. The columns contain: (1) the intrinsic pulsational velocity $v_{puls}$ of the bisector points; (2) the difference $v_{rad}-V_{\gamma}$, or in other words the observed pulsational velocity {\em
uncorrected for projection effects}, see the equation (2); (3-10) the X$_n$ coordinates of the eight
bisector points, or in other words the centers  of the bisectors in radial
velocity space, expressed as offsets from the gamma velocity (zero by
definition, in this system); (11-18) the Y$_n$ coordinates of the eight bisector
points,  expressed in relative flux (where 1 is the continuum and 0 full
absorption); and (19) the Full Width at Half Maximum of the synthetic LSD profile
(see text). The table is available in its entirety  on the electronic version of the Journal, and
at CDS. Only the first five rows are reported here, for guidance about its form
and contents.}
\label{tab-lib}
\begin{tabular}{cccccccccccccccccccccc}
\hline
 $v_{puls}$ &$v_{rad}-V_{\gamma}$*&	X1& ... & X8	&Y1& ... & Y8&	 FWHM \\
            km~s$^{-1}$&km~$s^{-1}$&km~s$^{-1}$ & ...&km~s$^{-1}$&(\%)&...&(\%)& $\AA$\\
\hline
 --45       &	--35.43&	0.32&	... & 3.53	&0.67&	... & 0.91&	 23.30 \\
  --44       &	--34.65&	0.37&	... & 3.49	&0.66&	... & 0.91&	 22.92 \\
 --43       &	--34.21&	0.42&	... & 4.24	&0.68&	... & 0.91&	 23.90 \\
 --42       &	--32.67&	0.29&	... & 3.56	&0.67&	... & 0.91&	 23.06 \\
 --41       &	--32.25&	0.27&	... & 3.62	&0.65&	... & 0.91&	 21.78 \\
\hline
\end{tabular}
\medskip \begin{flushleft} \emph{Notes.} * The accuracy in estimation of observed synthetic pulsational velocities ($v_{rad}-V_{\gamma}$) corresponds to the accuracy of finding minimum of the appropriate LSD profile (i.e. $\delta v_{obs}^{LSD}$), see the Section~\ref{sec-lsd}.
\end{flushleft}
\end{table*}

The grid is published in the electronic version of the Journal and
in CDS, and Table~\ref{tab-lib} provides an example of its contents.
To be able to use the grid, the set with the phase closest to the observed one needs to
be selected, and then the bisector's points need to be scaled to the actual
resolution of the observed spectra, by multiplying the X$_n$ coordinates of the
eight bisector points by $$\frac{FWHM_{LSD}}{FWHM_{obs}}$$ \noindent where
FWHM$_{LSD}$ is the Full-Width at Half Maximum of the synthetic bisector, reported
in the last column of Table~\ref{tab-lib}, while FWHM$_{obs}$ is the one obtained
for the observed spectra.

General suggestion is that to make more accurate determination of center-of-mass
velocity of radial pulsating Cepheids and RR Lyrae it is better to use
individual grids of synthetic bisectors. Since the line profile asymmetry depends on atmospheric parameters,
in ideal situation it is necessary to build the grid of synthetic spectrum with
$T_{eff}$, $log~g$ and $v_{mic}$  for a particular star, but the provide values can be used for stars that are not too different from RR Lyrae.


\label{lastpage}
\end{document}